\documentclass[prb,twocolumn,showpacs,amsmath,amssymb,footinbib]{revtex4}
\usepackage{graphicx,amsmath,bm}
\usepackage{ifthen}
\newboolean{Submittal}
\setboolean{Submittal}{true}	

\newcommand{\E}{\mbox{e}}
\newcommand{\I}{\mbox{i}}
\newcommand{\asin}{{\mathrm{Sin}^{-1}}}

\newcommand{\SpringConstant}{\ensuremath{K}}
\newcommand{\EquilMass}{\ensuremath{m_{\circ}}}
\newcommand{\MassPlus}{\ensuremath{m_+}}
\newcommand{\ImpurityMass}{\ensuremath{m_I}}
\newcommand{\UnitLength}{\ensuremath{a}}
\newcommand{\Disp}{\ensuremath{q}}	
\newcommand{\Mo}{\ensuremath{p}}	
\newcommand{\FFTDisp}{\ensuremath{Q}}
\newcommand{\MassDensity}{\ensuremath{\mu}}
\newcommand{\YoungsModulus}{\ensuremath{Y}}
\newcommand{\CrossSection}{\ensuremath{\sigma}}
\newcommand{\WaveNumber}{\ensuremath{k}}

\newcommand{\Wavelength}{\lambda}
\newcommand{\Concentration}{\ensuremath{c}}
\newcommand{\Length}{\ensuremath{L}}
\newcommand{\Viscosity}{\ensuremath{\eta}}
\newcommand{\GroupVel}{\ensuremath{v_{g}}}
\newcommand{\RelGVel}{\ensuremath{c_{s}}}

\newcommand{\Resistivity}{\rho}
\newcommand{\LocLength}{\xi}
\newcommand{\MFP}{\Lambda}		
\newcommand{\LocParam}{\ensuremath{\Gamma}}	
\newcommand{\Rmoment}{\ensuremath{M}}		
\newcommand{\Hmoment}{\ensuremath{H}}		
\newcommand{\Entropy}{\ensuremath{S}}
\newcommand{\Neff}{\ensuremath{n_\omega}}
\newcommand{\TCoeff}{\ensuremath{G}}		
\newcommand{\Energy}{\ensuremath{E_T}}		
\newcommand{\SiteE}{\ensuremath{E}}		
\newcommand{\Efluct}{\widetilde{E}}		
\newcommand{\ThermCond}{\kappa}			
\newcommand{\Temp}{\ensuremath{T}}		

\newcommand{\Etot}{\ensuremath{\Upsilon}}
\newcommand{\Nens}{\ensuremath{W}}

\newcommand{\SpectralDensity}{\ensuremath{u}}

\newcommand{\Texp}{\ensuremath{\delta}}		
\newcommand{\Error}{\epsilon}

\newcommand{\Ensemble}[1]{\left\langle #1\right\rangle}

\newcommand{\RNone}{\uppercase\expandafter{\romannumeral1}}
\newcommand{\RNtwo}{\uppercase\expandafter{\romannumeral2}}
\newcommand{\RNthree}{\uppercase\expandafter{\romannumeral3}}

\ifthenelse{\boolean{Submittal}}{%
\newcommand{\FigSource}[1]{}}{%
\newcommand{\FigSource}[1]{[#1]}}

\begin{document}
  \title{%
  	Energy transport along FPU-$\beta$ chains containing binary 
		isotopic disorder: {Z}ero temperature systems 
	}

  \author{K.A.\ Snyder}
	\affiliation{Materials and Construction Division, 
		National Institute of Standards and Technology}

  \author{T.R.\ Kirkpatrick}
	\affiliation{Institute for Physical Science and Technology and 
		Department of Physics, University of Maryland}

  \date{\today}


\begin{abstract}
Dissipation from harmonic energy eigenstates is used to 
characterize 
energy transport in binary isotopically disordered (BID)
Fermi-Pasta-Ulam (FPU-$\beta$) chains.
Using a continuum analog for the corresponding harmonic portion of 
the Hamiltonian,
the time-independent wave amplitude is calculated for a plane 
wave having wavelength $\Wavelength$ that is 
incident upon the disordered section, and the solution is mapped onto the 
discrete chain.
Due to Anderson localization, 
energy is initially localized near the incident end of the chain, and 
in the absence of anharmonicity the wave amplitude is stationary in time.
For sufficient anharmonicity, however, mode transitions lead to dissipation.
Energy transport along the chain 
is quantified using both the second moment $\Rmoment$ of the site 
energy, and the number of masses contributing to transport, which was 
estimated from the localization parameter.
Over the time scales studied, $\Rmoment$ increased linearly in 
time, yielding an effective transport coefficient $\TCoeff$.
At low and intermediate impurity concentration $\Concentration$, 
$\TCoeff(\Concentration)$ can be characterized by a competition 
between the rate of mode transitions and the time for energy to 
propagate a distance equal to the localization length $\LocLength$.
At the highest concentrations 
($1.6\le\Concentration\Wavelength\le 16.0$), 
there is significant mode transition suppression in BID systems, 
and the transport coefficient $\TCoeff(\Concentration)$ becomes
proportional to $\LocLength(c)$.
\end{abstract}

\maketitle

%
%

\section{Introduction}

Nonlinear binary disordered chains are a useful systems
for studying the essential characteristics of 
energy dissipation and transport in materials.
Although binary disorder is an idealized model,
it has practical application to a number of fields:
isotopic disorder effects line width broadening in spectroscopy;
\cite{Held97,Rohmfeld01,Widulle02}
the glass transition has been considered in terms of binary changes in 
elasticity parameters;\cite{Wagner92}
and 
isolated mechanical defects that occur in practice can lead 
to a variety of nonlinear effects.
\cite{Solodov02,Solodov04}

An interesting behavior of binary isotopic disorder (BID) occurs in discrete 
lattices, where the system undergoes a pure-disordered-pure 
transition as the impurity concentration varies from zero to one.
For harmonic one-dimensional chains composed of discrete elements, 
finite disorder destroys spatial invariance and gives rise to 
Anderson\cite{Anderson58} localization, characterized by 
spatially localized eigenstates. 
Given a discrete system in a localized energy eigenstate, 
the addition of anharmonicity will lead to interactions that 
create new modes that can
propagate through the system.  These propagating modes either will 
become localized or will undergo further mode transitions.

We are interested in the rate of energy dissipation from 
localized disturbances.
Previous studies of dissipation have used either a singular (one or very 
few elements) pulse or a Gaussian pulse for the initial displacement.
\cite{Bourbonnais90a,Bourbonnais90b,Sarmiento99,Yamada99,Rosas04}
Because 
neither is an eigenstate of the system, the singular pulse will spread 
and the Gaussian pulse will will 
propagate ballistically in both anharmonic and harmonic systems.
For Gaussian and `kicked' singular pulses, 
the initial behavior is ballistic.  In time, scattering and localization 
slow the rates of propagation and dispersion.

To eliminate the initial ballistic motion,
the system can be started in an
energy eigenstate of the correpsonding harmonic chain.
The middle section of the chain contains disorder, and there are 
``pure'' sections at both ends.  
Given the location of each impurity in the disordered section of the 
chain, a solution is found for the continuum analog, with the boundary 
condition of an incident plane wave with frequency $\omega$;
there is a reflective wave and a transmitted transmitted wave.
Because the continuum calculation is performed for the harmonic 
system, the solution is separable into spatial and temporal 
components.  The time dependence is sinusoidal, and the 
entire system has a constant temporal phase.
The continuum solution for the wave amplitude everywhere
is mapped onto the discrete chain, and becomes the initial displacement.

The advantage of the this approach is that in the 
absence of anharmonicity the wave amplitude is stationary.   
For the harmonic chain,
the wave remains localized indefinitely, and there is no 
energy transport.
The addition of anharmonicity will give rise to mode transitions that 
will de-localize the wave and lead to energy transport through the 
chain.   Instead of having a ballistic-diffusive transition, this 
initial condition leads to a localized-diffusive transition.

The remaining question is whether the energy will dissipate
in a diffusive manner.  In one-dimensional systems, mode interactions 
can lead to frequencies arbitrarily close to zero.  Because the localization 
length of these modes is proportional to $\omega^{-2}$
and the scattering cross 
section is proportional to $\omega^2$, these low frequency 
modes should propagate ballistically in a finite system.  
Because the energy content of these 
modes is proportional to $\omega^2$, however, it is unlikely that these 
ballistic modes contribute substantially to energy transport.

To study this energy transport,
the following numerical experiment will use the 
binary disordered Fermi-Pasta-Ulam\cite{Fermi55} chain with 
quartic spring potentials (FPU-$\beta$).
Starting from the initial energy eigenstate, 
numerical integration 
will be used to calculate the spatial distribution of energy as 
a function of time.
Based on a local concept of thermal transport,\cite{Wagner92} 
an effective transport coefficient will be sought from the 
second moment of the site energy, and the method is compared to the 
Helfand\cite{Helfand60} moments for thermal conductivity. 
The second moment will exhibit diffusive behavior, a fact that will 
be corroborated qualitatively from the number of masses that energy 
is distributed, which is estimated from the localization parameter.
\cite{Cretegny98,Piazza01}

%
%

\section{Numerical Experiment}

\subsection{FPU-$\beta$ Chain}

The FPU-$\beta$ chain is composed of masses interacting with nearest neighbors
through springs.
The Hamiltonian $H$ of a chain having
$N$ masses is function of the  mass
momenta $\Mo_i$ and the mass displacements $\Disp_i$ 
about their equilibrium position:
\begin{equation}
	H = \sum_i \frac{p_i^2}{2\,m_i} 
		+ \frac{\SpringConstant}{2} (\Disp_{i} - \Disp_{i-1})^2 
		+ \frac{\beta}{4} (\Disp_{i} - \Disp_{i-1})^4
	\label{eqn:Hamiltonian}
\end{equation}
The harmonic spring force coefficient $\SpringConstant$
is set equal to one, and 
in the absence of impurities, each mass has the same value $\EquilMass$=1.
The site energy $\SiteE_i$ is the sum of the kinetic energy plus 
one-half of the neighboring spring potential energies.  The total 
energy $\Energy$ is the sum of the site energies.

Whenever possible, the results are expressed in dimensionless 
units through the use of appropriate scaling factors.
Time is scaled by the 
natural frequency $\omega_o$ of a single harmonic oscillator:
\begin{equation}
	\omega_o = \left(\frac{\SpringConstant}{\EquilMass}\right)^{1/2} 
\end{equation}
Lengths are scaled by the equilibrium mass separation distance
$a$.

The time-dependent behavior was determined by numerical 
integration using 
the sixth-order Yoshida \cite{Yoshida90} symplectic 
integration algorithm.   Specifically, best results were 
obtained from the ``Solution A'' coefficients 
(see Table~1 in Ref.~\onlinecite{Yoshida90}).
For the systems studied here, 
the time step $\Delta t$ was approximately 1/200 the period of 
the initial mode.  
This time step was consistent with that used  
elsewhere,\cite{Livi85,Bourbonnais90a} 
was chosen as a compromise between speed and accuracy,  
and the results were insensitive to two-fold changes in $\Delta t$.
Using this time step, the energy fluctuations were
always less then 0.2~\% (See Appendix).

\subsection{Semi-Infinite Approximation}

In most cases, low frequency waves will
propagate down the chain and reach the far end.  If the wave
is reflected, it could affect the accuracy of the transport
calculation.   To mitigate this effect, 10~\% of 
the masses at the far end were given a viscous force 
$F_{vis}$:
\begin{equation}
	F_{vis} = - \Viscosity\, \dot{\Disp}
\end{equation}
This approach has been used elsewhere to 
achieve a similar effect.\cite{Khomeriki04}
For these calculations, a viscosity $\Viscosity$ of 0.2 was 
sufficient to eliminate the effects of reflection.

The viscous damping, combined with the accurate time integration,
simplified the task of identifying finite size effects.  As the 
system length decreased and the localization length increased, 
the likelihood of a considerable amount of energy reaching the 
far end of the system increased.  Fortunately, this occurrence was 
easily identified by changes in the system total energy.

\subsection{Initial Displacement}

The initial condition was a stationary 
state exhibiting Anderson localization for the harmonic 
component of the Hamiltonian.
This initial condition was chosen so that for $\beta$=0 there is 
no net energy transport,
and these systems could be used as a test to confirm the 
accuracy of the model and the numerical integration.

The continuum Kronig-Penney liquid model \cite{Kronig31,Ziman79} is used to 
estimate the initial displacement.  
For the harmonic FPU chain, the analogous continuum system is an 
elastic rod having mass density 
$\MassDensity = \EquilMass/\UnitLength$ 
and Youngs modulus $\YoungsModulus=\SpringConstant\UnitLength$.
Between the impurities, a longitudinal wave 
$\psi(x,t|\omega)$ with frequency $\omega$ 
will propagate with
velocity $c_l = \sqrt{\YoungsModulus/\MassDensity}$.
A harmonic oscillator impurity, approximated by a point defect, 
located at $x^\prime$ will give rise to a 
reactive force due to the impurity impedance $Z$ to a wave with
frequency $\omega$: \cite{Morse68}
\begin{equation}
   \left[
	\MassDensity\frac{\partial^2}{\partial t^2}
	- \YoungsModulus \frac{\partial^2}{\partial x^2} = 
	-Z(\omega)\,\delta(x-x^\prime) \frac{\partial}{\partial t}
	   \right]\,\psi(x,t|\omega)
  \label{eqn:continuum}
\end{equation}
For a harmonic system such as this, one can assume
a sinusoidal solution with frequency $\omega$ 
($\psi=\phi(x|\omega)\,\E^{-\I\omega t}$):
\begin{equation}
   \left[\frac{\partial^2}{\partial x^2} + \zeta^2  
   	= \frac{-\I\omega}{\YoungsModulus}\,
		Z(\omega)\,\delta(x-x^\prime)
	\right] \phi(x|\omega)
   \label{eqn:Helmholtz}
\end{equation}
where $\zeta=\omega/c_l$.

Equation~(\ref{eqn:Helmholtz})
is solved for a system having 
impurities at integer 
locations with probability 
$\Concentration$.
The boundary conditions are a unit amplitude incident 
wave and a reflected wave at one end and 
only a transmitted wave at the other end.
The initial displacement for the physical 
system of masses and springs is taken from the real component of the solution 
$\phi(x|\omega)$.

\subsection{Impurity Cross Section}

To put some of the results in a familiar context,
it will be useful to characterize an
impurity by its cross section to the original wave.
The cross section $\CrossSection$ of an individual impurity can 
be expressed as a function of the impurity impedance $Z$:
\cite{Morse68}
\begin{equation}
	\sigma = \frac{\left|Z(\omega)\right|^2}{\left|Z(\omega)\right|^2
			+ 4\SpringConstant\EquilMass}
\end{equation}
The isotopic impurities consist of a constant mass $\MassPlus$ added 
to the existing mass $\EquilMass$; $\MassPlus$ may be either positive 
or negative.  
(The impurity mass $\ImpurityMass = \EquilMass+\MassPlus$.) 
The impedance $Z$ of a mass impurity in a continuum system 
\cite{Morse68} is modified by $\RelGVel$, which is the ratio 
of the group velocity ($\GroupVel = \partial\omega/\partial\WaveNumber$)
to the longitudinal velocity $c_l$:\cite{Snyder04}
\begin{equation}
	Z(\omega) = -\I \omega \MassPlus / \RelGVel
\end{equation}
For the discrete chain, 
$\RelGVel = \cos(\WaveNumber\UnitLength/2)$, where 
$\WaveNumber$ is the wavenumber ($2\pi/\lambda$) and 
$\Wavelength$ is the displacement wavelength.

\subsection{Continuum-Discrete Mapping}

The boundary condition for the chain is zero displacement at each
end.
Because the continuum solution $\phi(x=0,\Length|\omega)$ will (with almost 
certainty) not equal zero 
at ($x=0,\Length$),
a method is needed for adjusting the continuum solution to accommodate the 
constraints of the discrete chain.
The continuum solution is first mapped to the 
discrete chain with no modification, and then, starting at the end and 
searching along the chain,
the end is relocated at the mass having the smallest oscillation
amplitude. 
The end is relocated to this mass,
and its displacement is fixed at zero.
The process is repeated 
at the opposite end of the chain.

To facilitate this task, a suitable initial 
displacement wavelength $\Wavelength$
is needed to ensure that some mass has an equilibrium 
displacement acceptably close to zero.
If the displacement wavelength is an integer multiple of
the equilibrium mass displacement $\UnitLength$, 
the oscillation amplitude of masses along each 
successive wavelength remains constant,
until the next impurity changes the 
phase.

For this experiment, a nominal displacement (equal to an integer 
multiple of the unit spacing) is chosen first.  The working wavelength
is the nominal wavelength minus 0.2 unit spacings.  With this 
wavelength, the displacement closest to zero repeats every 5 
wavelengths, with 9 nodal points occurring at different fractions of 
the spacing $a$.  
Probabilistically, 
this reduces the minimum mass oscillation amplitude by nearly an order 
of magnitude over an integer wavelength.

\begin{figure}
\ifthenelse{\boolean{Submittal}}{%
  \includegraphics[width=\columnwidth]{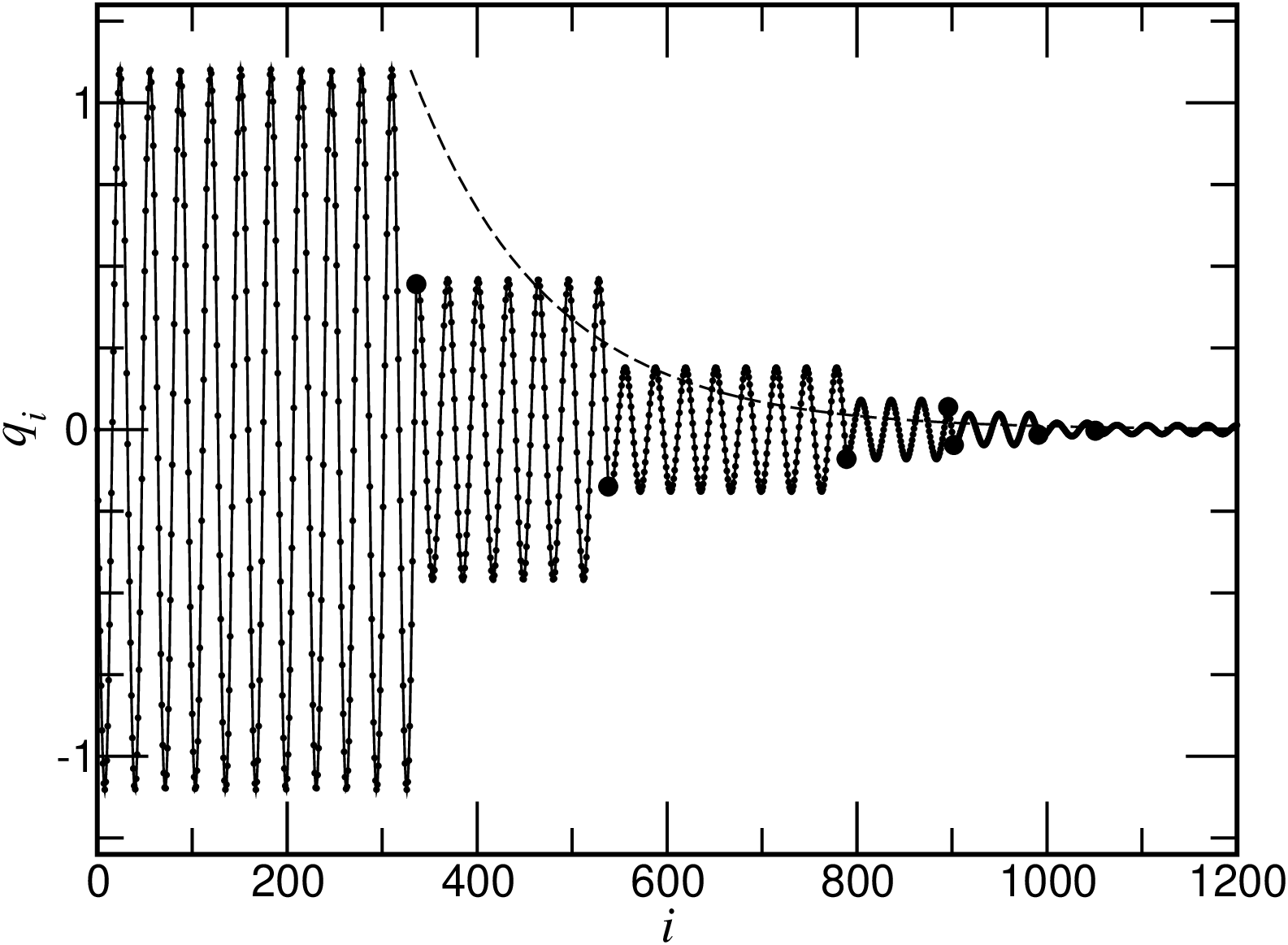}}{%
  \includegraphics[width=\columnwidth]{GRAPHS/InitCheck}}
  \caption{Initial displacements $\Disp_i$ for a particular system having
  	wavlength 31.8, 
	impurity concentration 0.01, and 
	impurity scattering cross section 0.5.  
	Mass displacement are denoted by small circles and 
	the impurity locations are denoted by large circles.
	The first impurity is located at $i=\iota$, and 
	the dashed line is proportional to $\E^{-(i-\iota)/\LocLength}$.
  \FigSource{InitCheck}}
  \label{InitCheck:fig}
\end{figure}

An example input displacement is shown in Fig.~\ref{InitCheck:fig} for 
a system with length 10\ 000, 
displacement wavelength 31.8, 
impurity cross section 0.5, 
and impurity concentration 0.01.  
The small dots in the figure represent the displacement of the masses.  
The larger filled circles denote the location of impurities.  The effect of 
the impurities is to change both the the displacement amplitude and phase.
Also shown in the figure is a dashed line 
that is proportional to $\E^{-(i-\iota)/\LocLength}$, where $\iota$ 
is the location of the first impurity.
Although this particular initial condition would 
suggest that the displacement amplitudes decrease 
monotonically, that is not always the case.

%
%

\subsection{Dense Systems}


For BID systems, the localization length is not a monotonic function 
of impurity concentration.
Rather, the localization length 
has a minimum with respect to 
impurity concentration; at higher concentrations, the 
system returns to a ``pure'' system.
The localization length $\LocLength$ of BID systems at dilute 
impurity concentrations, such that $\Concentration\Wavelength \ll 1$, 
can be calculated from the resistivity scaling law:
\cite{Anderson80}
\begin{equation}
	\LocLength_{\Concentration\rightarrow 0}^{-1} 
		= c \ln\left(1+\Resistivity\right)
\end{equation}
The single impurity 
resistitivity $\Resistivity$ is related to the single impurity 
cross section $\CrossSection$:\cite{Landauer70}
\begin{equation}
	\Resistivity = \frac{\CrossSection}{1-\CrossSection}
\end{equation}

As the impurity concentration increases,
the system approaches a homogeneous system.  
In the limit $\Concentration\rightarrow 1$, the system is again ordered, and 
the localization length diverges:
\begin{equation}
	\LocLength_{\Concentration\rightarrow 1}^{-1} 
		= (1-c)\ln(1+\Resistivity^\prime)
\end{equation}
The quantity $\Resistivity^\prime$ characterizes a system having an
equilibrium mass $(\EquilMass+\MassPlus)$, an impurity mass
$\EquilMass$, and the system oscillates at the same frequency $\omega$:
\cite{Snyder04}
\begin{subequations}
  \label{highc:eqn}
  \begin{eqnarray}
     \WaveNumber^\prime      & = & \frac{2}{a}\,
	  \asin\left[\frac{\omega}{2}\,
			\sqrt{\frac{\EquilMass+\MassPlus}{\SpringConstant}}\,
		\right] \label{kprime:eqn}\\
     \RelGVel^\prime & = & \cos\left(\WaveNumber^\prime \UnitLength/2\right) \\
     \Resistivity^\prime & = & \frac{(-\MassPlus\omega/\RelGVel^\prime)^2}{%
                4\SpringConstant(\EquilMass+\MassPlus)}
  \end{eqnarray}
\end{subequations}
The relation for $\WaveNumber^\prime$ in Eq.~(\ref{kprime:eqn}) is an 
improvement over 
the low frequency approximation given previously.\cite{Snyder04}
Because the localization length is analogous to conductivity, and 
the systems described by 
$\LocLength_{\Concentration\rightarrow 0}$ and 
$\LocLength_{\Concentration\rightarrow 1}$ 
occur independently and in parallel,
the localization length over all values of impurity concentration 
can be approximated by a sum of the two:\cite{Snyder04}
\begin{equation}
	\LocLength = 
		\LocLength_{\Concentration\rightarrow 0} +
		\LocLength_{\Concentration\rightarrow 1}
  \label{xi:eqn}
\end{equation}

\begin{figure}
\ifthenelse{\boolean{Submittal}}{%
  \includegraphics[width=\columnwidth]{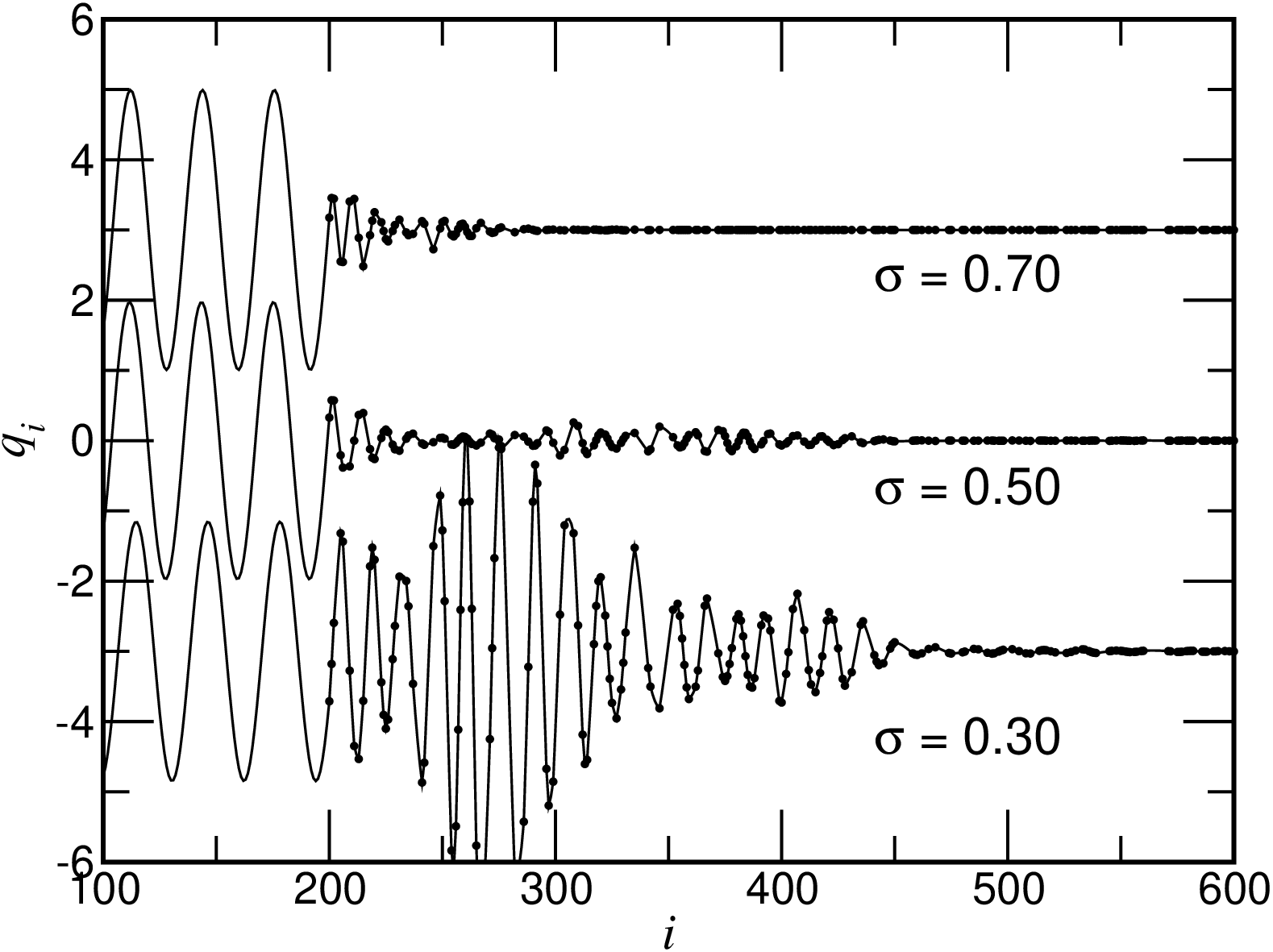}}{%
  \includegraphics[width=\columnwidth]{GRAPHS/InitCheck500}}
  \caption{Initial displacements $\Disp_i$ for a particular system having
  	wavlength 31.8, 
	impurity concentration 0.50, and 
	all three impurity scattering cross sections.
	Mass displacement are denoted by line, and 
	the impurity locations are denoted by circles.
	The systems were shifted horizontally so that the first 
	impurity is located at $i=$200.
  \FigSource{InitCheck500}}
  \label{InitCheck500:fig}
\end{figure}

When the concentration of the impurities increases to
$\Concentration\Wavelength>1$, the energy eigenstate becomes more 
complicated than the dilute impurity example shown in 
Fig.~\ref{InitCheck:fig}.  Example initial conditions for the 
highest impurity concentration 
considered in this experiment ($\Concentration = 0.5$) are shown 
in Fig.~\ref{InitCheck500:fig} for three scattering cross 
sections.  At these high concentrations, although the 
frequency remains constant everywhere, 
the wave structure in the disordered region is complex.


\subsection{Parameter Space}

\begin{figure}
\ifthenelse{\boolean{Submittal}}{%
   \includegraphics[width=\columnwidth]{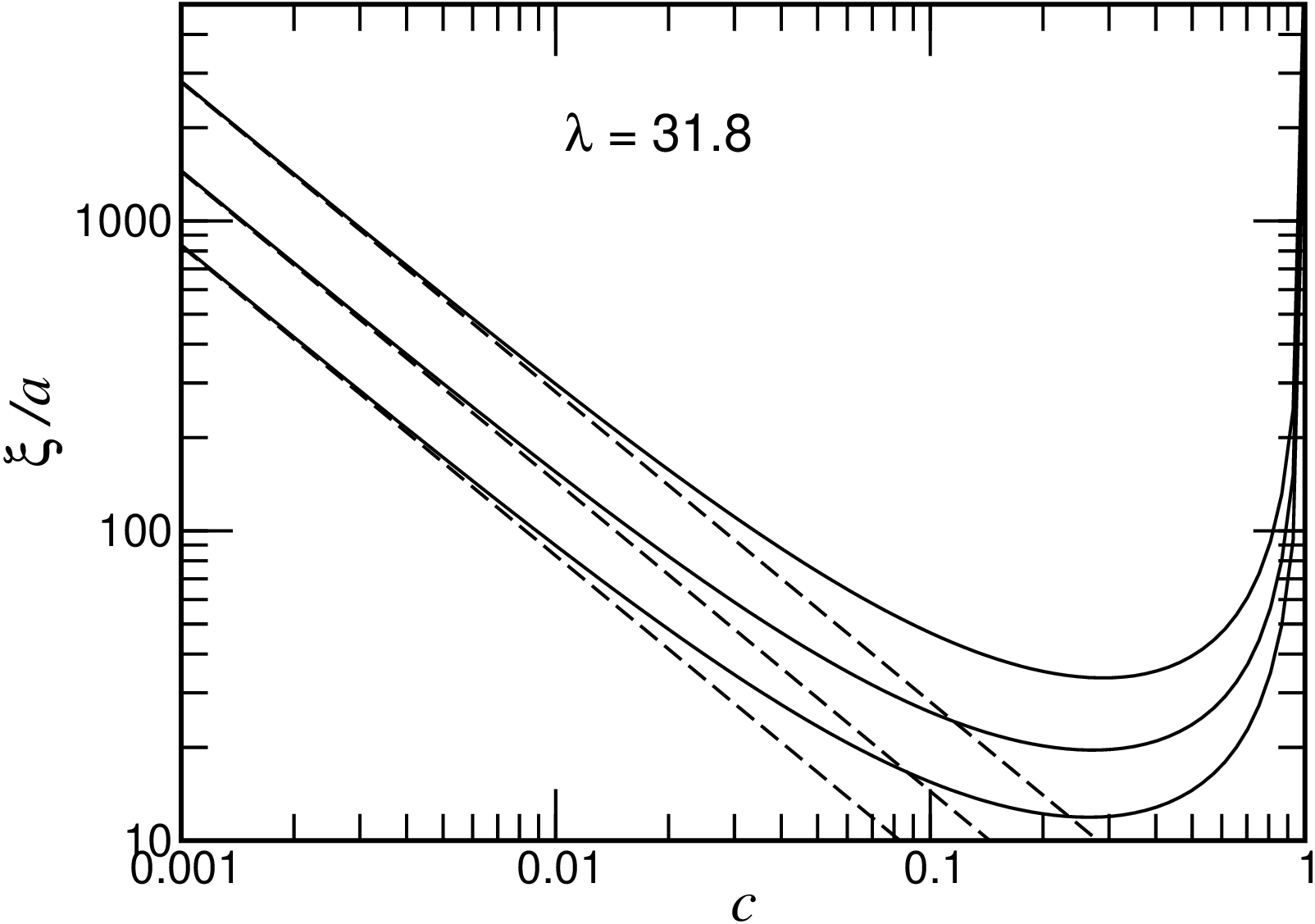}}{%
   \includegraphics[width=\columnwidth]{GRAPHS/xi_318}}
	\caption{Localization length $\LocLength$ as a function of 
		impurity concentration $\Concentration$ for systems having 
		displacement wavelength $\Wavelength =$ 31.8.  The three 
		solid curves, from upper to lower, 
		are for $\CrossSection =$ 0.30, 0.50, and 0.70.
		The dashed lines are the dilute limit localization length
		$\LocLength_{\Concentration\rightarrow 0}$.
	\FigSource{xi\_318}}
	\label{xi_318:fig}
\end{figure}

For the nonlinear systems considered, the incident wave has
wavelength $\Wavelength = $31.8,  
and the anharmonicity parameter $\beta = $1.  
For all the systems from which energy transport is measured, the 
added mass $\MassPlus$ will have one of three values:
6.605, 10.089, and 15.412.  With respect to a $\Wavelength=31.8$ 
displacement in a harmonic system, these impurity masses 
have scattering cross sections 0.30, 0.50, and 0.70, respectively.

The initial displacement is a localized mode with 
localization length 
$\LocLength$.  The localization lengths calculated from 
Eq.~(\ref{xi:eqn}) for $\Wavelength=31.8$ and 
the three impurity masses mentioned are shown  
in Fig.~\ref{xi_318:fig} as a function of impurity concentration 
$\Concentration$.
The dilute limit result 
$\LocLength_{\Concentration\rightarrow 0}$ is shown as a dashed line,
and begins to depart from $\LocLength$ 
near $\Concentration\Wavelength\sim 1$.

Figure~\ref{xi_318:fig} also reveals the utility of choosing
a nominal displacement wavelength $\Wavelength=32$.  
For $\Concentration\Wavelength > 1$, the addition of impurities 
has a nonlinear effect on localization length.  
Naturally, one would like to see whether this nonlinear 
behavior has any effect on energy transport.
For shorter displacement wavelengths, the deviation between 
$\LocLength$ and 
$\LocLength_{\Concentration\rightarrow 0}$  would not 
occur until proportionately higher impurity concentrations.  Alternatively, 
using a 
longer wavelength would reduce the rate of mode transitions 
considerably, requiring excessively long computational times.

%
%

\section{Thermal Conduction}

There are a number of formal methods for calculating thermal conductivity.
Thermostats at the boundaries can generate a steady-state flux from 
which the thermal conductivity is calculated using Fourier's law.
\cite{Payton67}
The Green-Kubo method\cite{Evans90} is 
based on time integrals of thermal fluxes for a 
system in thermal equilibrium.
The Evans NEMD thermal conductivity algorithm\cite{Evans82,Evans90}
applies a heat field and calculates an averaged heat flux.
The Helfand\cite{Helfand60} moments of the execess energy fluctuations 
are calculated for systems in thermal equilibrium.
In each case, the system is some state of steady-state or 
thermal equilibrium, which does not apply here.

Elsewhere, the second moment of the site energy has been used to 
characterize pulse dissipation.
\cite{Bourbonnais90a,Bourbonnais90b,Sarmiento99,Yamada99,Rosas04}
This approach bears a qualitative resemblance to the method of 
Helfand, which is based on mean squared fluctuations of excess 
energy.
Here, a brief summary of the Helfand moments for thermal conductivity is used 
to motivate the relevance of the second moment of the site energy 
in nonequilibrium chains at zero temperature.

The thermal conductivity of a collection of freely moving 
particles in thermal equilibrium 
can be determined from energy fluctuations.  The 
energy fluctuation $\Efluct_i$ for the $i$-th particle is the 
difference between the instantaneous site energy $\SiteE_i$ and the 
ensemble averaged value $\Ensemble{\SiteE_i}$:
\begin{equation}
	\Efluct = \SiteE_i - \Ensemble{\SiteE_i}
\end{equation}
If the energy fluctuation is conserved, and the energy flux has a linear 
dependence on $\nabla\Efluct$, the quantity $\Efluct_i(x,t)$ will 
satisfy the diffusion equation.  For the boundary conditions that the 
initial value $\Efluct_i(x,0)$ is localized about $x_{i0}$, and that 
$\Efluct_i(\pm\infty,t) = 0$, the solution for $\Efluct_i(x,t)$ is 
Gaussian, and the measure of spread is the second moment of $\Efluct_i$:
\begin{equation}
	\int (x-x_{i0})^2\,\Efluct(x,t)\,dx \sim 2\Efluct_i(x,0)
		\ThermCond t
\end{equation}
where $\ThermCond$ is the thermal conductivity.

As $\Efluct_i$ represents the fluctuation for a single particle, a 
bulk expression requires an ensemble integral of the second moment.
Making no assumption about the independence of particle energies,
the thermal conductivity can be calculated from a double sum over 
particle positions:
\cite{Helfand60}
\begin{equation}
	\Hmoment^p = 
	\left\langle\,\sum_{i,j}\,(x_i-x_{j0})^2\,\Efluct_i(x,t)\,\Efluct_j(x,0)
		\right\rangle \sim 2\ThermCond t
	\label{Hp:eqn}
\end{equation}
Replacing conservation of momentum with conservation of energy yields 
an equivalent alternative expression:
\cite{Helfand60}
\begin{equation}
	\Hmoment^e = 
	\left\langle
	  \left[\sum_i(x_i\Efluct_i - x_{i0}\Efluct_{i0})\right]^2\right\rangle
		\sim 2 \ThermCond t
	\label{He:eqn}
\end{equation}
Equations~(\ref{Hp:eqn}) and (\ref{He:eqn}) are the Helfand moments for 
calculating the thermal conductivity of a bath of particles.

There are a number of differences between fluctuations in a bath of 
particles and 
energy propagation along a discrete chain.
Equations~(\ref{Hp:eqn}) and (\ref{He:eqn}) characterize a bath of freely 
moving particles.  By contrast, in the FPU chain, energy moves, but the 
masses are, more or less, stationary.  This is not entirely problematic, 
however, because one can still evaluate the energy that is at $x_i$.  
The problem can be changed to one in which the energy is evaluated 
at specific points.  
In this way, 
the role of energy is analogous to concentration in the 
evaluation of self-diffusion.

Another important distinction is that Eqs.~(\ref{Hp:eqn}) and (\ref{He:eqn})
are functions of the energy fluctuations in a system in equilibrium 
at temperature $\Temp$.  By contrast, the pulse moving through the FPU 
chain is a system that is not in equilibrium, and the portion of the chain 
farthest from the initial disturbance is initially at zero temperature.
In principle, after very long times, the chain would eventually reach 
equilibrium, with the energy distributed over all the masses.  Because 
the conceptual problem of interest is a semi-infinite chain, the 
equilibrium energy $\langle \SiteE_i\rangle$ would approach zero.
Under this assumption, the fluctuation energy $\Efluct$ of the 
bath problem becomes the site energy $\SiteE_i$ of the non-equilibrium 
chain problem.

By analogy to Eqs.~(\ref{Hp:eqn}) and (\ref{He:eqn}),
the energy transport in the 
FPU-$\beta$ chain will be characterized by the second moment of the 
energy.   Assuming that the initial pulse occupies a small portion of 
the entire system, 
a useful measure is the second moment about zero:
\begin{equation}
	\frac{\sum_i r_i^2 \SiteE_i}{\sum \SiteE_i} \sim 2 G t
\end{equation}
The position $r_i = ia$ is the equilibrium location of the $i$-th mass.
The quantity $G$ is an effective transport coefficient that is neither 
self-diffusion nor thermal conductivity.  To eliminate the effects of 
fluctuations, the initial value is subtracted from the 
subsequent values.  In addition, the equation is generalized to
allow for arbitrary powers of $\SiteE_i$:
\begin{equation}
	\Rmoment_n(t) = \left\langle
	   \frac{\sum_i r_i^2 \SiteE_i^n}{\sum \SiteE_i^n}  
		- \Rmoment_n(0)\right\rangle
	\sim 2 G_n t
\end{equation}
These definitions are similar to those used by 
Fr{\"o}lich et al.,\cite{Frohlich86}
and are consistent with the local concept of thermal transport of 
Wagner et al.\cite{Wagner92}


\begin{figure}
\ifthenelse{\boolean{Submittal}}{%
   \includegraphics[width=\columnwidth]{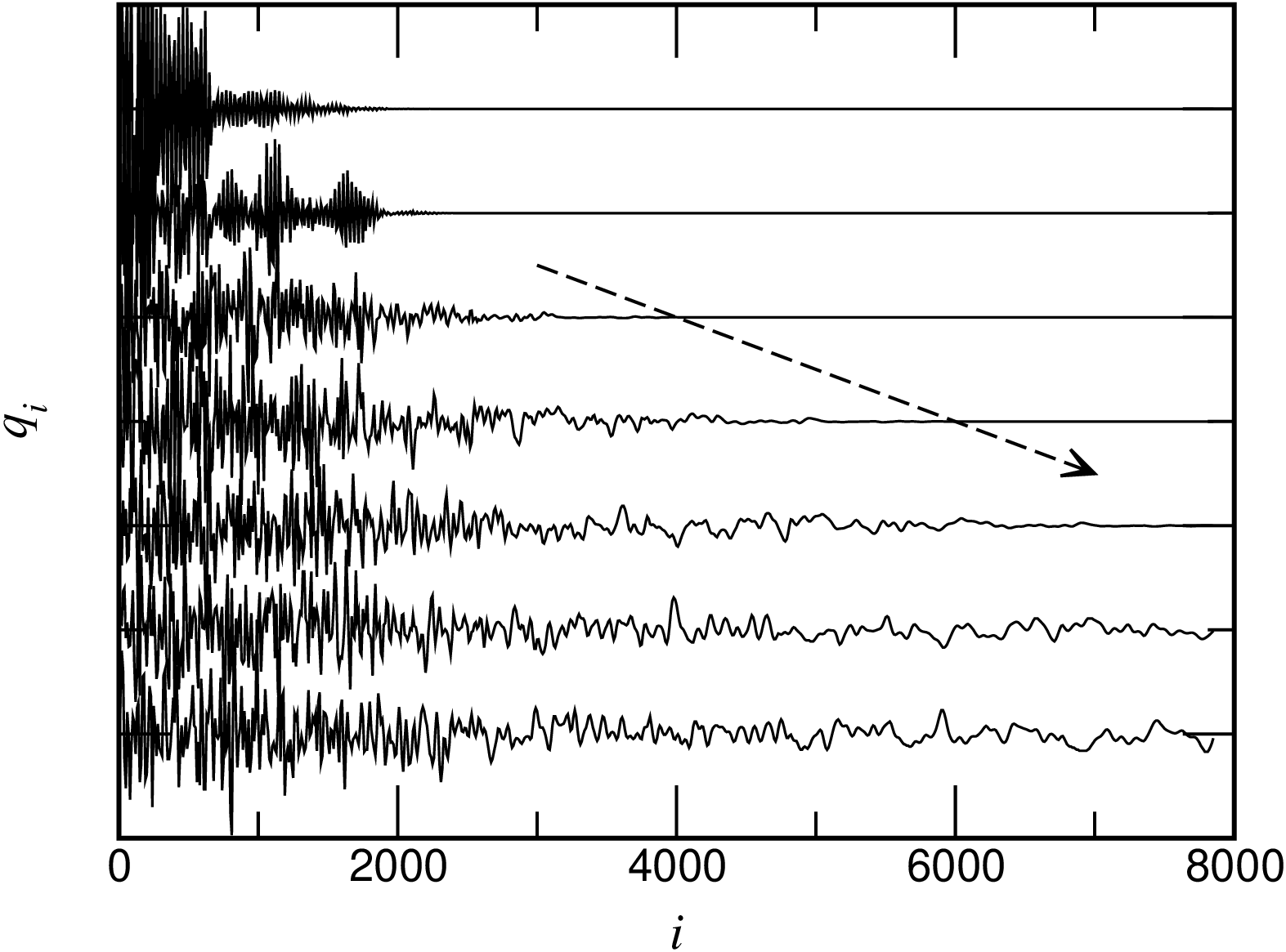}}{%
   \includegraphics[width=\columnwidth]{GRAPHS/DEMO/pulse}}
	\caption{Wave displacement $\Disp_i$ along a chain at various times.
		Chain length is 8000, added mass $\MassPlus$ is 10.089, 
		and impurity concentration is 0.010.
		Each curve represents a time difference of 2000, and 
		is offset by a value of one for demonstration purposes.
		Dashed arror denotes ballistic propagation.
	\FigSource{DEMO/pulse}}
	\label{pulse:fig}
\end{figure}

A comparison among $\Rmoment_1$, $\Rmoment_2$, 
$\Hmoment^p$, and $\Hmoment^e$ is made from the 
early response of a system of length 8000 
that is initially in a localized mode.  The
wavelength is 31.8, the impurity mass is 11.089, the impurity 
concentration is 0.010, and $\beta =$ 1.
The displacement $\Disp$ along the system is shown 
in Fig.~\ref{pulse:fig} at time intervals of $\Delta\omega_o t =$ 2000.
(The curves are offset vertically from one another, by a distance 
$\UnitLength$, for comparison 
purposes.)  The data in the figure show that
long wavelength displacements move virtually ballistically along the 
chain (parallel to dashed arrow) 
while the higher frequency displacements propagate a considerably 
shorter distance over the same time interval.

\begin{figure}
\ifthenelse{\boolean{Submittal}}{%
   \includegraphics[width=\columnwidth]{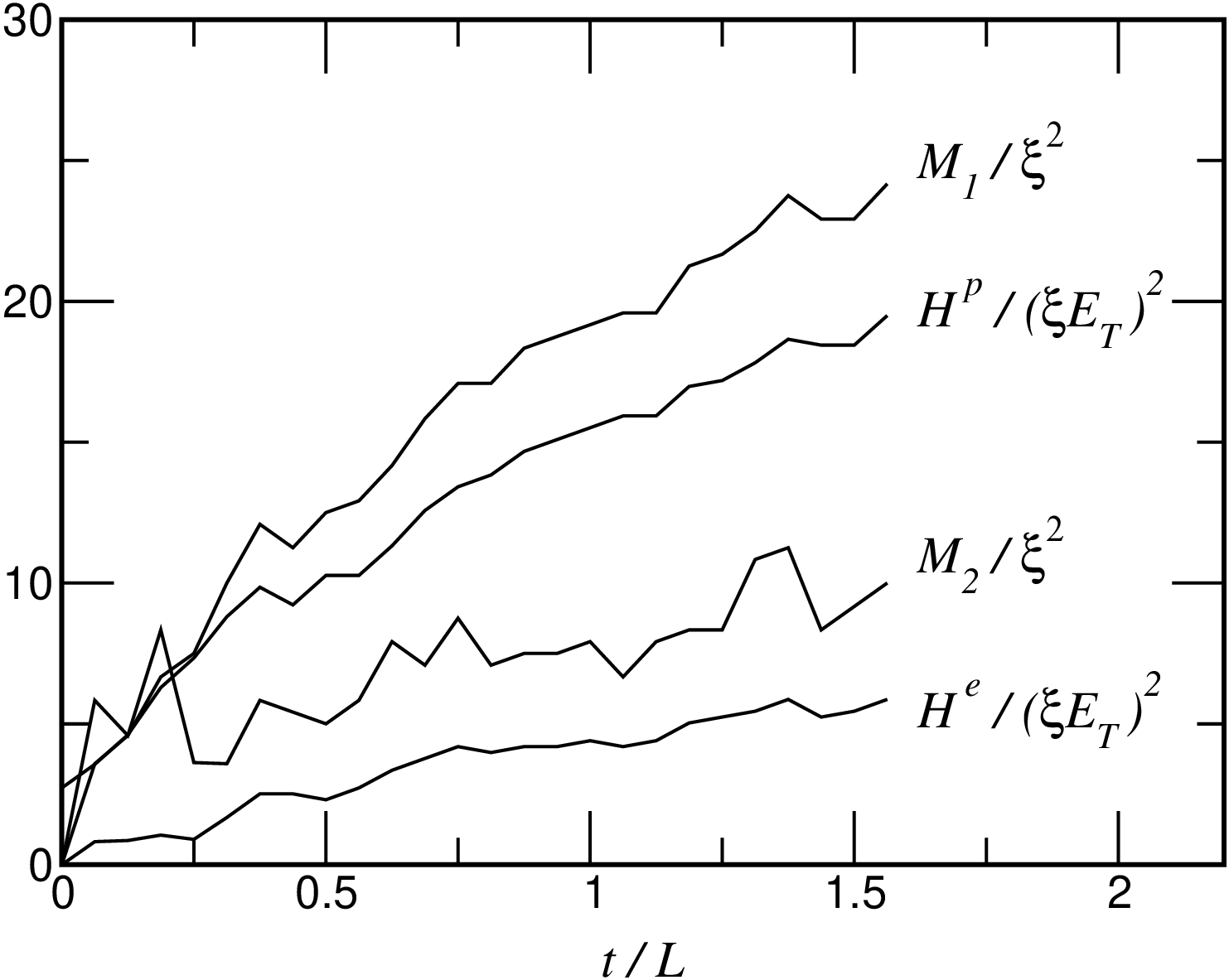}}{%
   \includegraphics[width=\columnwidth]{GRAPHS/DEMO/R2M2}}
	\caption{Moments $\Rmoment_1$, $\Rmoment_2$, 
		$\Hmoment^p$, $\Hmoment^e$ as a function of time for the 
		system shown in Fig.~\protect{\ref{pulse:fig}}.  
		Quantities 
		are normalized, using localization length $\LocLength$
		and total energy $E_T$, to make the values dimensionless.
	\FigSource{DEMO/R2M2}}
	\label{R2M2:fig}
\end{figure}

At regular time intervals, $\Rmoment_1$ and $\Rmoment_2$ 
are calculated, 
along with $\Hmoment^p$ and $\Hmoment^e$ (assuming that $\Efluct=E_i$ and 
$\langle \SiteE_i\rangle = 0$). 
The results of the calculations are shown in Fig.~\ref{R2M2:fig}.
All values are normalized by the 
initial localization length ($\LocLength\approx 155$).  
$\Hmoment^p$ and $\Hmoment^e$ 
are also normalized by the total energy $\Energy$ to be on the same
scale as $\Rmoment_1$ and $\Rmoment_2$.  
The two pairs of equations correspond well 
to one another.  Generally, $\Rmoment_1$ and $\Rmoment_2$ 
are greater in value than the 
corresponding $\Hmoment^p$ and $\Hmoment^e$.  At the shortest times,
$\Rmoment_1$ and $\Hmoment^p$ give the nearly the same value.

$\Hmoment^p$ and $\Hmoment^e$ are equivalent descriptions of 
thermal conductivity.  Therefore, the 
difference between $\Hmoment^p$ and $\Hmoment^e$ for the case of 
an initially localized pulse demonstrates that thermal conductivity 
for these systems is not well-defined.  As a result, the second moments
$\Rmoment_n$ characterize some effective, yet undefined, transport 
coefficient.

These curves are also instructive in pointing out the distinction between 
pulse propagation and energy propagation.  From Fig.~\ref{pulse:fig} it is 
clear that low frequency waves propagate ballistically through these 
systems, starting from near $t=0$.  
By contrast, Fig.~\ref{R2M2:fig} shows no ballistic behavior, suggesting 
that the vast majority 
of the energy is in the higher frequency modes located within the initially 
localized region.  Moreover, the moments shown in Fig.~\ref{R2M2:fig} all
continue to increase after there has been sufficient time for the 
low frequency ballistic modes to reach the far end.


\section{Initial Condition}

\begin{figure}
\ifthenelse{\boolean{Submittal}}{%
   \includegraphics[width=\columnwidth]{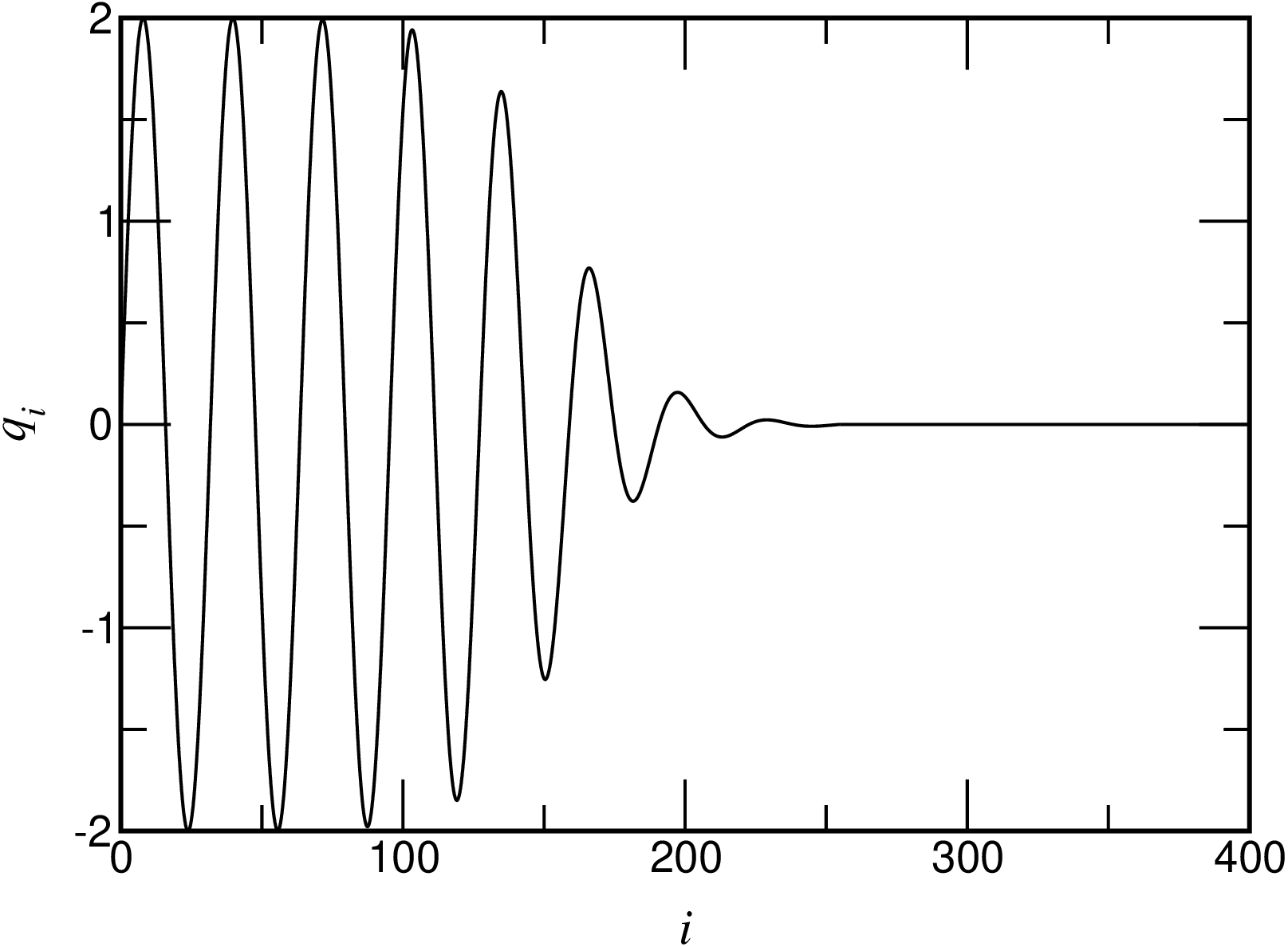}}{%
   \includegraphics[width=\columnwidth]{GRAPHS/sinusoid}}
  \caption{Displacement $\Disp_i$ for sinusoid initial 
	condition with $\Wavelength=31.8$.
  \FigSource{sinusoid}}
  \label{sinusoid:fig}
\end{figure}

The initial condition, composed of a localized mode, has the advantage 
of spatial stability.
In the absence of anharmonicity, the initial wave will oscillate with a 
periodic amplitude that is constant in time.

Alternatively, 
this experiment could have been performed using either 
an instantaneous impulse or a short
sinusoidal pulse like that shown in Fig.~\ref{sinusoid:fig}.  Both of these 
initial conditions, however, have drawbacks.  Impulses will impart 
relatively little energy to the system unless the impulse is large.
The sinusoidal wave, not being localized by the impurities, will 
immediately begin propagating ballistically along the chain.  If the chain 
had only harmonic interactions, the wave would eventually settle into 
a localized state.  In a nonlinear chain 
there will be immediate competition between localization of the initial 
wavelength and anharmonic effects that would lead to mode transitions that 
would then experience the same competition.

\begin{figure}
\ifthenelse{\boolean{Submittal}}{%
   \includegraphics[width=\columnwidth]{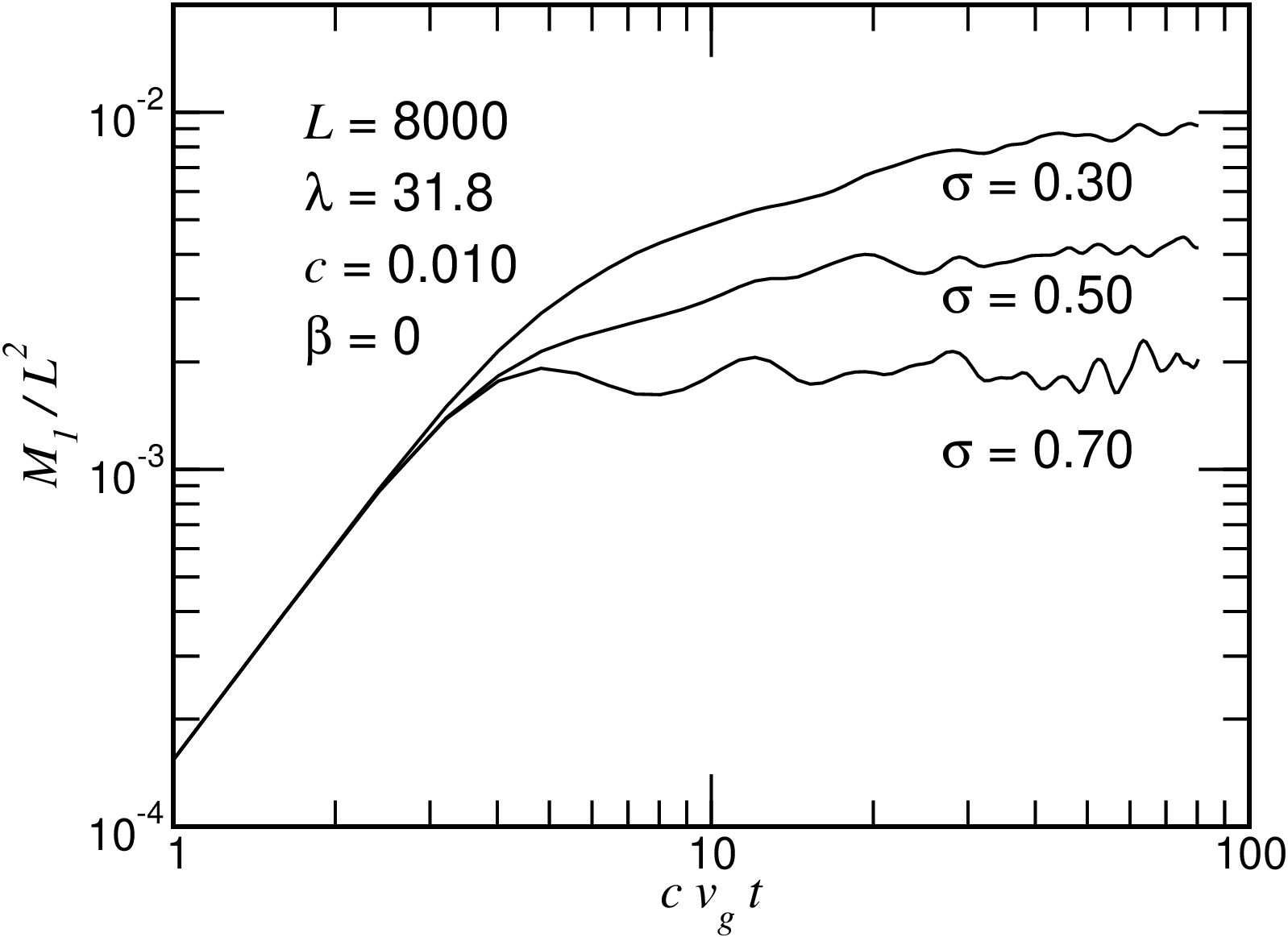}}{%
   \includegraphics[width=\columnwidth]{GRAPHS/SINUSOID/p8k_010_10}}
  \caption{$\Rmoment_1$ as a function of time $t$ for 
	systems with anharmonic parameter $\beta=0$ and 
	impurity concentration $\Concentration=0.010$ for 
	systems having length $\Length=8000$.
  \FigSource{SINUSOID/p8k\_010\_10}}
  \label{p8k_010_10:fig}
\end{figure}

The time required for the sinusoidal pulse to become localized
in a harmonic chain is shown in 
Fig.~\ref{p8k_010_10:fig} for the $\Wavelength=31.8$ pulse shown 
in Fig.~\ref{sinusoid:fig}.  The pulse is located at one end of the 
system, the envelope is the hyperbolic tangent function, 
and the initial velocity is zero.
The harmonic chains have mass impurities located at a site with probability 
$\Concentration$.  Three different different values for $\ImpurityMass$ 
are considered, corresponding to 
$\CrossSection$ = 0.30, 0.50, and 0.70.  The data in 
Fig.~\ref{p8k_010_10:fig} suggest that somewhere between 10 and 100 
impurities, depending on the impurity cross section, 
must be encountered before the sinusoidal wave becomes localized.
The localization length for these three cross sections are 
298, 155, and 90, respectively, so the transient period is approximately
ten times the localization length.
For the lowest impurity concentrations considered in this experiment, 
this transition length
would have made the required system length prohibitively long.


\section{Equipartition}

The rate of both mode transitions and spatial energy equipartition 
will influence the response of the systems.  The mode 
transition rate for similar chains configured as small hoops 
initially excited in a single eigenmode studied previously\cite{Snyder99}  
suggests that impurities initially hasten the decay of energy in the 
excited mode.  Over long times, however, the rate diminished because 
the energy became localized at the impurities.  Because the impurities 
were heavier than the background, the oscillations at the impurities was 
smaller, reducing the rate of energy loss because of the quadratic 
dependence of amplitude.

The systems studied here, however, have energy initially localized at 
one end, with the energy already localized at the impurities.
Once transitions start to occur, 
the new modes, which are not localized over the same length scale, 
will propagate and both spontaneously decay and collide with impurities.

\subsection{Localization Parameter}

As the mode transitions occur, energy will propagate along the 
chain, redistributing energy.  
As the energy becomes more evenly distributed among the masses, 
the energy becomes less localized.
A measure of how uniformly the energy is distributed among $N$
masses is the localization parameter $\LocParam$:
\cite{Cretegny98,Piazza01}
\begin{equation}
	\LocParam = N\,\left\langle
		\frac{\sum^N E_i^2}{\left(\sum^N\,E_i\right)^2}
	  \right\rangle
	\label{LocParam:eqn}
\end{equation}
The value of $\LocParam$ is a minimum for ergodic behavior and 
increases as the degree of localization increases.

The localization parameter can be used to estimate the number of 
masses over which energy is distributed.  The maximum value of 
$\LocParam$ is $N$, when all the energy is localized at 
one mass.  At long time, $\LocParam$ approaches a constant,
$\LocParam_\infty = \LocParam(t\rightarrow\infty)$,
that only depends on the value of $\beta$.\cite{Cretegny97}
For the FPU-$\beta$ system, with $\beta=1$, the equilibrium value 
$\LocParam_\infty$ is approximately 1.8.  
Thus, $\LocParam_\infty/\LocParam$
is approximately equal to the portion of the chain over which energy 
is distributed.

\subsection{Participating Modes}

Another useful measure of ergodicity is the number of 
harmonic modes contributing to the overall energy at a mass.  The systems 
studied will be initially excited in one mode 
(in frequency space).  The time required for the system to 
excite the maximum number of modes should correspond to the 
time required for the system to become ergodic.

The energy in a particular mode $E_\omega$ is estimated from 
the harmonic approximation involving the Fourier 
transformed (FT) displacement $\FFTDisp_i(\omega)$ 
and momentum $P_i(\omega)$ of mass $m_i$:
\begin{equation}
  E_{i,\omega} = \frac{1}{2}\left(m_i \omega^2 \FFTDisp_i^2 
		+ \frac{P_i^2}{m_i}\right)
\end{equation}
The energy is then normalized using the total number of frequency modes
$N_\omega$ considered in the FT:
\begin{equation}
	e_{i,\omega} = E_{i,\omega} / \sum_\omega^{N_\omega} E_{i,\omega}
\end{equation}
These normalized energies are a measure of energy entropy $\Entropy_i$ at 
mass $m_i$:
\cite{Brillouin56,Livi85,Cretegny98,Luca02}
\begin{equation}
	\Entropy_i = -\sum_\omega \, e_{i,\omega}\,\ln(e_{i,\omega})
\end{equation}
If all the energy is in a single frequency mode, $\Entropy$ equals $0$.  
If the energy is distributed evenly among all frequency modes, 
$\Entropy = \ln N_\omega$.

The alternative expression for the energy entropy is 
$\exp\left(\Entropy\right)$.  This quantity is the equivalent 
number of modes contributing to the overall entropy if the energy 
is uniformly distributed among those modes.
To make comparisons among results using different 
values for $N_\omega$, results are expressed as the fraction of 
modes $\Neff$:
\begin{equation}
	\Neff^{(i)} = \frac{\exp\left(S_i\right)}{N_\omega}
\end{equation}
Periodically, $\Neff^{(i)}$ is calculated at various masses
and the reported value, $\Neff(t)$, is the average of these values.

\subsection{Hoop Example}

The localization parameter $\LocParam$ and the fraction of participating 
modes $\Neff$ were developed to study systems in which the energy is 
initially distributed throughout.  For the systems studied here, 
the initial energy is intentionally localized at one end of the system.
If such a system was divided into two equal halves, the values for 
$\LocParam$ and $\Neff$ in one half would be very different from the 
ones calculated for the other half of the system.  Nonetheless, the 
parameters do have utility for these systems.




One application is the study of behavior within a short section of chain.
If the section of chain is quasi-localized (very few modes
are present, oscillating with nearly constant amplitude), mode transitions
are the primary mechanism for inducing energy transport.  A 
hoop (periodic boundary conditions), initially excited in one 
wave number mode, could be used to study the behavior of a similar section
within a much longer section that is itself quasi-localized.  The 
time-dependent behavior of the localization parameter $\LocParam$ and 
the relative number of participating modes $\Neff$ would characterize 
energy redistribution, with respect to both space and mode frequency.

A brief numerical calculation of $\LocParam$ and $\Neff$ is made to 
study the effect.
The initial condition is a BID hoop having length 318,
wavelength 31.8, and $\beta=1$.  
This initial condition differs from 
a localized state.  In a disordered system,
a single $\WaveNumber$ mode will excite 
multiple $\omega$ modes, accelerating 
the initial rate of mode transitions.
Nevertheless, the results illuminate general behavior for nonlinear 
BID systems.

\begin{figure}
\ifthenelse{\boolean{Submittal}}{%
   \includegraphics[width=\columnwidth]{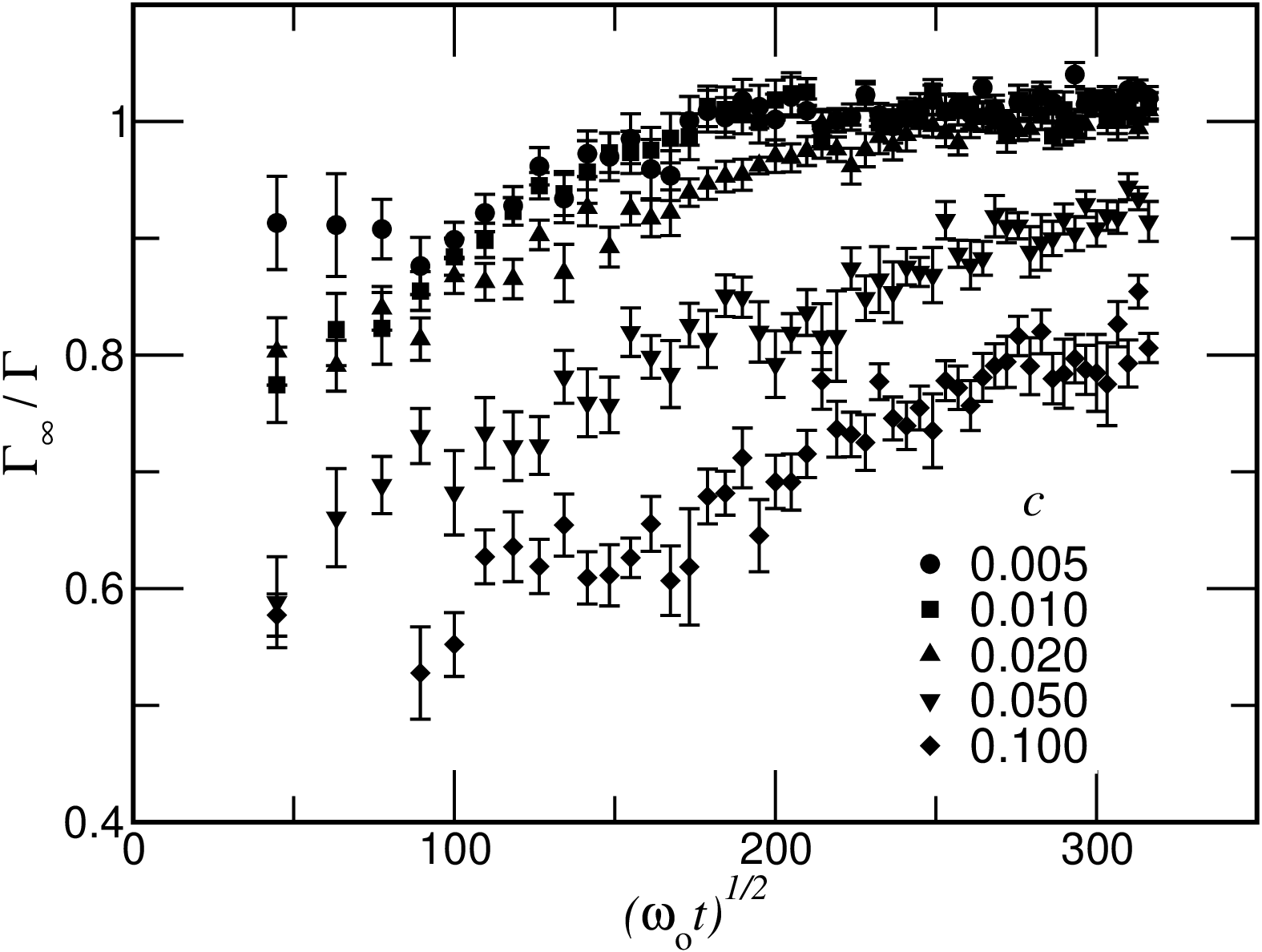}}{%
   \includegraphics[width=\columnwidth]{GRAPHS/ERGODIC/erg636_Lt_50}}
  \caption{Localization parameter $\LocParam$ as a function of time 
	$t$ for a periodic system with length 636, 
	initial wavelength 31.8, impurity cross section 0.5, 
	and anharmonicity 1.0.
  \FigSource{ERGODIC/erg636\_Lt\_50}}
  \label{erg636_Lt_50:fig}
\end{figure}

\begin{figure}
\ifthenelse{\boolean{Submittal}}{%
   \includegraphics[width=\columnwidth]{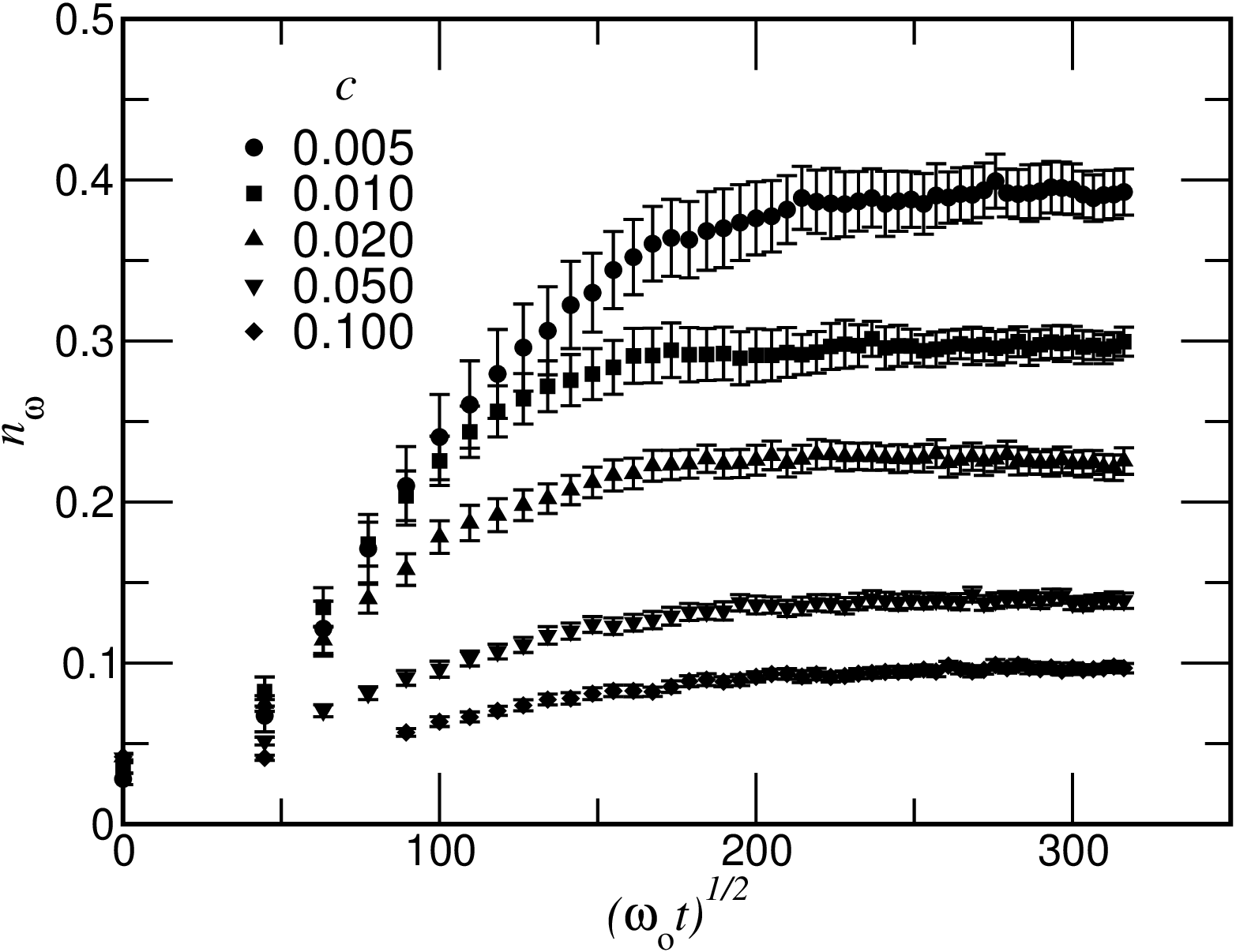}}{%
   \includegraphics[width=\columnwidth]{GRAPHS/ERGODIC/erg636_Neff_50}}
  \caption{Fraction of participating modes $\Neff$ as a function of time 
	$t$ for a periodic system with length 636, 
	initial wavelength 31.8, impurity cross section 0.5, 
	and anharmonicity 1.0.
  \FigSource{ERGODIC/erg636\_Neff\_50}}
  \label{erg636_Neff_50:fig}
\end{figure}


The effects of impurity concentration on energy distribution and on
mode production are shown in Figs.~\ref{erg636_Lt_50:fig}
and \ref{erg636_Neff_50:fig}.   For dilute 
systems ($\Concentration\Wavelength < 1$), the time required for 
energy to be distributed among all the masses is nearly a constant.
For concentrated systems ($\Concentration\Wavelength > 1$), the spatial 
redistribution of energy is slowed dramatically.  

This behavior is consistent with the data for $\Neff$ in 
Fig.~\ref{erg636_Neff_50:fig}.  The rate 
that new modes are produced in dilute systems is a constant, until 
$\Neff$ approaches its asymptotic value.  
At concentrations for which $\Concentration\Wavelength > 1$, however,
the initial rate of mode production does not reach the common initial 
rate.
The asymptotic value for 
$\Neff$ decreases with increasing impurity concentration. 
Therefore, concentrated systems produce fewer modes at a slower 
rate than less concentrated systems.

The decreasing asymptotic value for $\Neff$ with increasing impurity 
concentration indicates that modes are suppressed significantly in 
concentrated systems.  This is consistent with the effect of impurities
on the spectral density $\SpectralDensity$ of 
BID systems.
The presence of impurities 
forces zeroes in $\SpectralDensity(\omega)$,
\cite{Saxon49,Luttinger51}
thereby constraining mode transitions.  
Moreover, as impurity concentration increases, the spectral density in 
the interval 
$\left[4\SpringConstant/\ImpurityMass 
	\leq\omega^2 
	\leq 4\SpringConstant/\EquilMass\right]$  
(assuming $\ImpurityMass > \EquilMass$)
becomes increasingly suppressed, eventually containing 
isolated delta functions.\cite{Hori68,Payton68}  
If the frequency of the initial displacement is in the 
interval $\left[0\leq\omega^2\leq 4\SpringConstant/\ImpurityMass\right]$,
the mode will be in a continuous portion of $\SpectralDensity(\omega)$ for
all values of impurity concentration.  As the impurity concentration 
increases, new modes will be generated more slowly, and in 
proximity to the originial frequency.


\section{Length and Time Scales}

At the lowest impurity concentrations, the chain is composed of 
long segments of homogeneous nonlinear chain.  As new modes are 
generated, these modes are not localized and begin to propagate ballistically 
along the chain.  The wave will continue to propagate until 
it scatters from an impurity or undergoes a spontaneous 
transition.  A spontaneous transition along the homogeneous portion 
of the chain is unlikely to occur in the time required to span the 
distance between impurities.  It is more likely that the impurities 
will initiate scattering and mode transitions.  
The mean free path $\MFP=(\Concentration\CrossSection)^{-1}$ characterizes 
the effective transport coefficient $D(\omega)$:
\begin{equation}
	D(\omega) = \GroupVel\MFP 
\end{equation}
Based on this assumption, thermal dissipation coefficient should 
be proportional to $\Concentration^{-1}$ at the lowest concentrations.

As the impurity concentration increases, the localization length decreases 
and the wave experiences considerably more wave interference, 
and transport is more diffusive because 
any propagation is now limited by localization.
The relevant time scale is the time $t_{\LocLength}$ 
required for energy to diffuse a distance comparable to the localization 
length $\LocLength$:
\begin{equation}
	D(\omega) = \frac{\LocLength^2}{t_{\LocLength}}
	\label{Dxi:eqn}
\end{equation}
When this occurs, the thermal transport coefficient should become 
proportional to $\Concentration^{-2}$.

%
%

\section{Results}

The primary experimental parameter 
was impurity concentration $\Concentration$.
The impurity concentration lower limit was constrained by computing 
resources.  
A previous study of these systems revealed that 
approximately 32 impurities are required in a chain to ensure reliable 
ensemble statistics.\cite{Snyder04}
The upper concentration limit was 0.5.  Above this concentration,
the behavior is indistiguishable from a complimentary study of the pure system 
having mass $\ImpurityMass$ and impurity mass $\EquilMass$.

Some of the results include estimates of uncertainty for 
quantities calculated from ensemble averages.  For a calculation
performed on an ensemble of $\Nens$ systems, there is a population 
standard deviation $s$ and an average value.  For this study, 
the average value is the meaningful quantity.
The uncertainty
in the reported average value is $s/\sqrt{\Nens}$, and is referred to 
here as the standard deviation in the mean (SDM).


There are two special cases within the parameter space that require 
special consideration.  The $\Rmoment_n$ data for the nonlinear systems 
are only meaningful when the $\Rmoment_n$ data for the corresponding 
harmonic system is a constant.  This must be true for all cases, 
especially at concentrations for which 
$\Concentration\Wavelength \gg 1$.

\begin{figure}
\ifthenelse{\boolean{Submittal}}{%
   \includegraphics[width=\columnwidth]{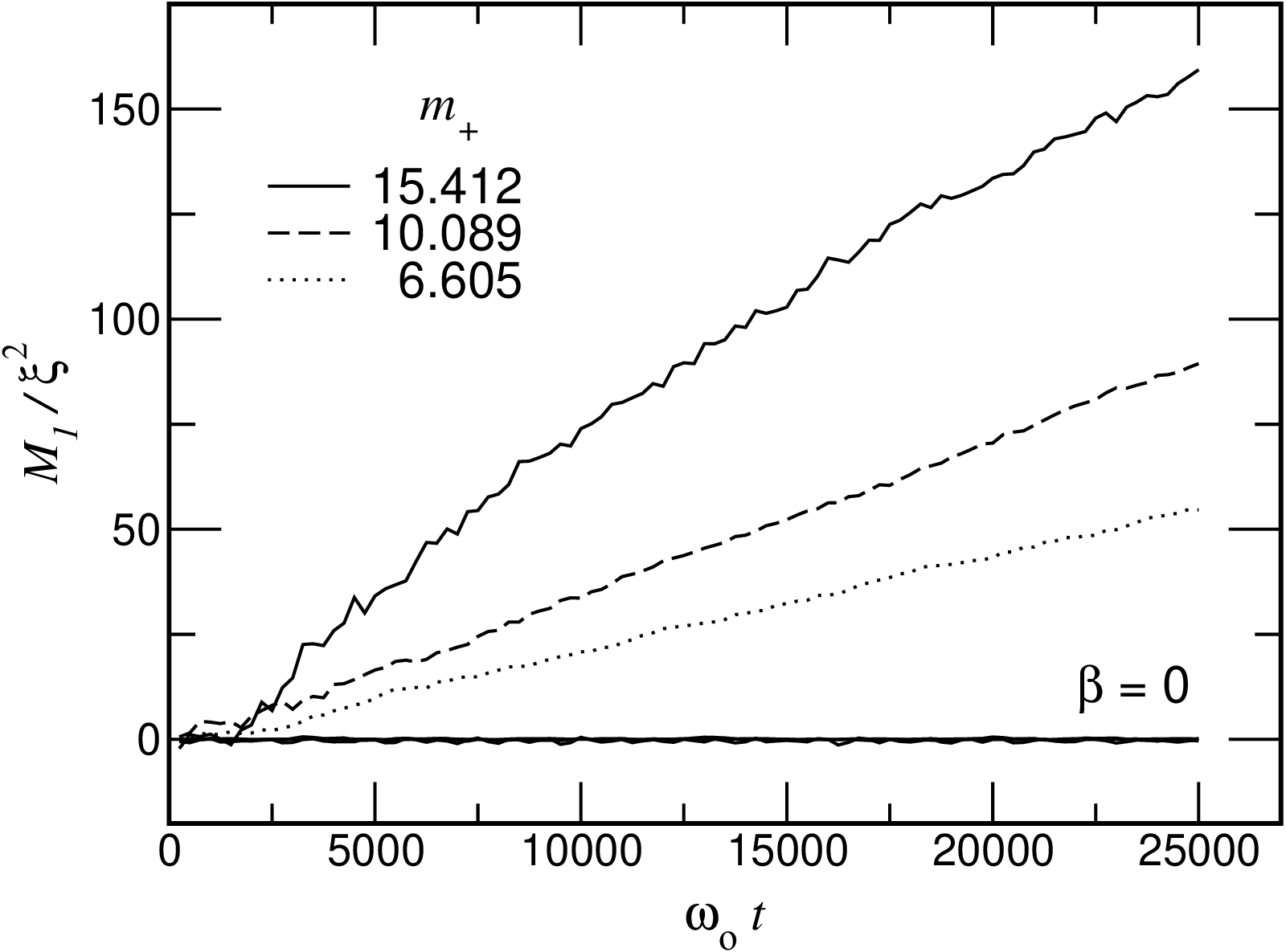}}{%
   \includegraphics[width=\columnwidth]{GRAPHS/h10_500}}
	\caption{The ratio $\Rmoment_1/\LocLength$ as  a function of time 
		for the systems having 
		impurity concentration 0.500
		and length 16\ 000.
		The $\beta=1$ data have positive slopes and all the 
		$\beta=0$ data fall on top of one another.
	\FigSource{h10\_500}}
	\label{harmonic:fig}
\end{figure}

In theory, $\Rmoment_n$ for harmonic systems would be a constant
for all time.  In practice, the mapping of the continuum system onto the 
discrete lattice, and the relocation of the ends, introduced a small amount
of instability that led to small fluctuations in $\Rmoment_n$.  These 
fluctuations, however, were far smaller than the changes 
for the anharmonic systems.  

As a brief example, Fig.~\ref{harmonic:fig} is a plot of $\Rmoment_1$ 
as a function of time for systems having having impurity 
concentration 0.500 and length 16\ 000.  In the figure, the 
$\beta = 1$ data appear as lines having positive slope.  The 
harmonic $\beta=0$ data for all three impurity masses lie 
upon one another near $\Rmoment_1 = 0$.  Figure~\ref{harmonic:fig}
is doubly instructive.  It demonstrates that the oscillations 
for the $\beta=0$ data are negligible, even for the systems for which 
$\Rmoment_1$ has the smallest values.  Moreover, the harmonic 
data remain stable in concentrated systems:
$\Concentration\Wavelength \gg 1$.

\subsection{Time Exponent}

\begin{table}
  \caption{\label{texp:table}Interval from which $\Texp_n$ was 
	calculated for systems having length $\Length$.}
  \begin{ruledtabular}
    \begin{tabular}{crcr@{\ $ < \omega_o\, t < $}lc}
	\hspace{1cm}
	& \multicolumn{1}{c}{$\Length$} 
	& 	
	& \multicolumn{2}{c}{Interval} 
	& \hspace{1cm}  \\
			\hline
	\ &16000	& \ &   5000 	&  20000 & \ \\
	\ &32000	& \ &   10000 	&  25000 & \ \\
	\ &64000	& \ &  20000 	&  50000 & \ \\
	\ &96000	& \ &  20000	&  80000 & \ \\
    \end{tabular}
  \end{ruledtabular}
\end{table}

A quantitative characterization of $\Rmoment_n$ is made by assuming a 
power-law dependence on time $t$:
\begin{equation}
	\Rmoment_n = 2\TCoeff_n t^{\Texp_n}
\end{equation}
The first task is to determine the value of the exponent so as to 
distinguish among ballistic, diffusive, or sub-diffusive transport.
The calculation of $\Texp_n$ is affected by initial transients, and the 
details of the analysis are given in the Appendix.
The time interval from which the values of $\Texp_n$ were calculated 
were constrained by the initial transients at small times and total 
energy conservation at long times.  The specific intervals are 
shown in Table~\ref{texp:table} for each system length.

\begin{figure}
\ifthenelse{\boolean{Submittal}}{%
   \includegraphics[width=\columnwidth]{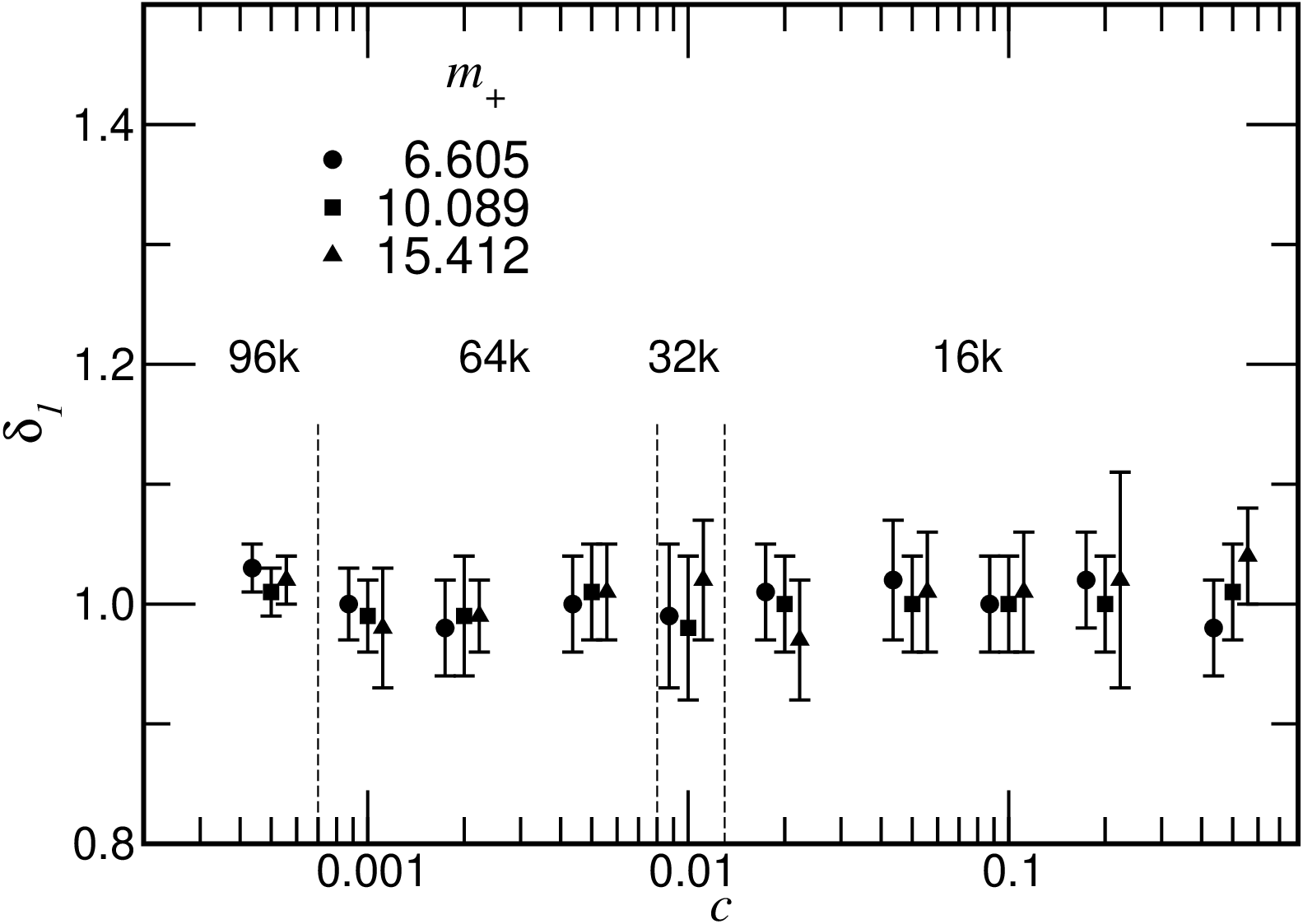}}{%
   \includegraphics[width=\columnwidth]{GRAPHS/texp}}
  \caption{The time exponent $\Texp_1$ for $\TCoeff_1$ 
	as a function of impurity concentration 
	$\Concentration$ for impurity cross sections 
	$\CrossSection =$ 0.30, 0.50, 0.70.
	Numbers between vertical dashed lines denote $N$
	(k=1000).
  \FigSource{texp}}
  \label{Texp:fig}
\end{figure}

The results of the analyses for $\Texp_1$ are summarized in 
Fig.~\ref{Texp:fig} for all the systems.   (In the figure, 
the symbols at a particular concentration 
are displaced horizontally to distinguish individual 
error bars representing the SDM.)   With only one exception, the estimated 
values for $\Texp_1$ are within one SDM of 1.
Therefore, subsequent analysis of $\Rmoment_1$ is based on 
linear model with respect to time.

\begin{figure}
\ifthenelse{\boolean{Submittal}}{%
   \includegraphics[width=\columnwidth]{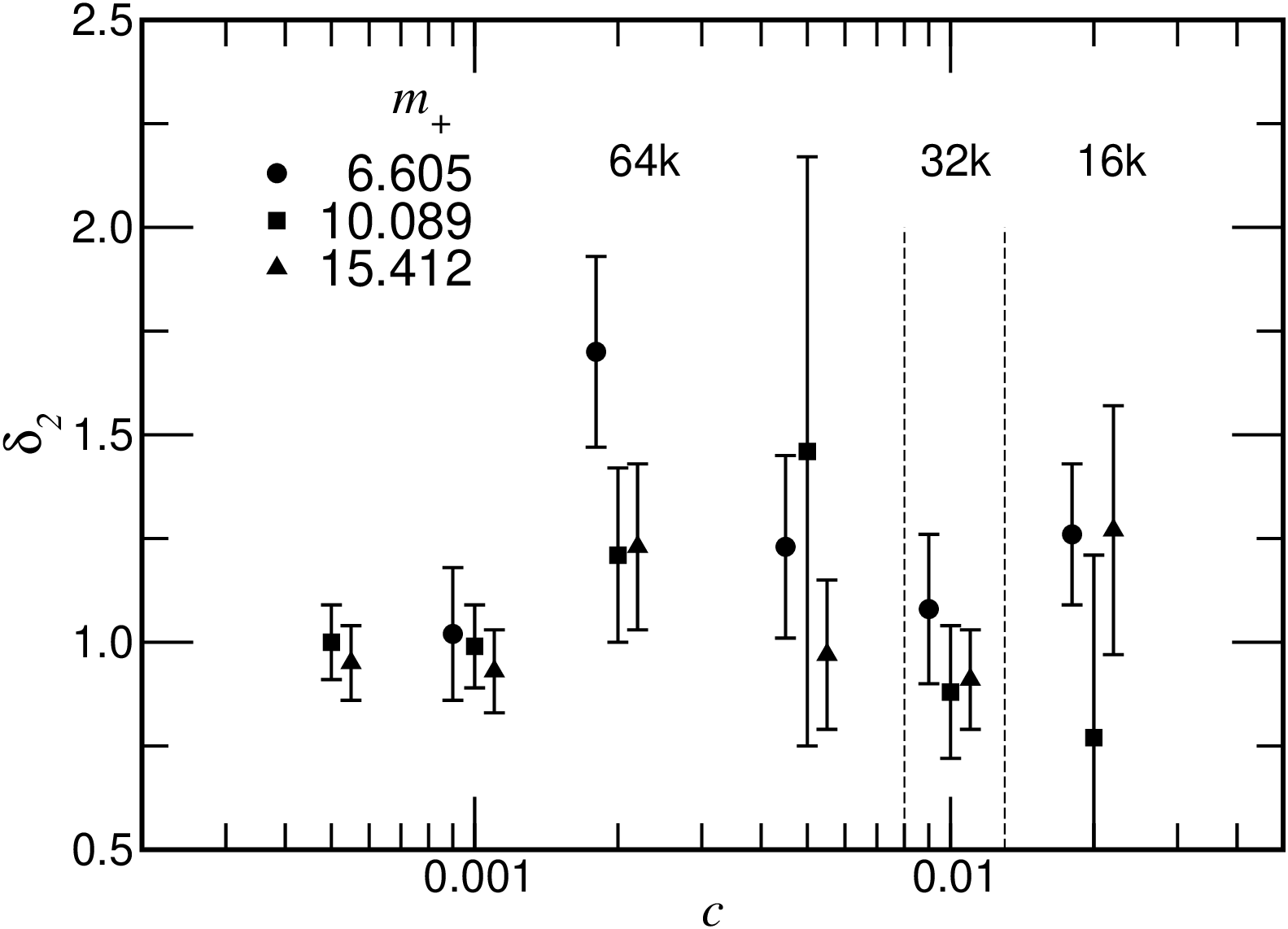}}{%
   \includegraphics[width=\columnwidth]{GRAPHS/texpM2}}
  \caption{The time exponent $\Texp_2$ for $\TCoeff_2$ 
	as a function of impurity concentration 
	$\Concentration$ for impurity cross sections 
	$\CrossSection =$ 0.30, 0.50, 0.70.
	Numbers between vertical dashed lines denote system size 
	used (k=1000).
  \FigSource{texpM2}}
  \label{TexpM2:fig}
\end{figure}

Results of the analysis for $\Texp_2$ are shown in Fig.~\ref{TexpM2:fig} 
for the same systems.  Although the values of $\Texp_2$ are near 1 for 
most systems, there is considerably more variability than for $\Texp_1$.
Moreover, it was difficult to establish a precise value for 
$\Texp_2$ at higher concentrations.
As a result, subsequent analysis is confined to $\Rmoment_1$. 

According to Li et al.,\cite{Li03} if the mean squared displacement of 
a particle is proportional to $t^\alpha$ with $(0\leq\alpha\le 2)$, the 
thermal conductivity can be expressed as a function of the system 
size $L$: $\kappa \sim L^b$ with $b = 2 - 2/\alpha$.  Therefore, 
a system obeying normal diffusion implies normal heat conduction 
obeying Fourier's law $(b=0)$.

\subsection{Transport Coefficient}

Because $\Texp_1$ was shown to be within one SDM of 1, the determination
of 2$\TCoeff_1$ is based on the assumption of a linear relationship 
between $\Rmoment_1$ and $t$
over the same intervals shown in 
Table~\ref{texp:table}.  
Although $\Rmoment_1$ was determined by linear regression with a linear 
model, calculating the uncertainty in $\Rmoment_1$ required a slightly 
more involved analysis; details are give in the Appendix.  

\begin{figure}
\ifthenelse{\boolean{Submittal}}{%
   \includegraphics[width=\columnwidth]{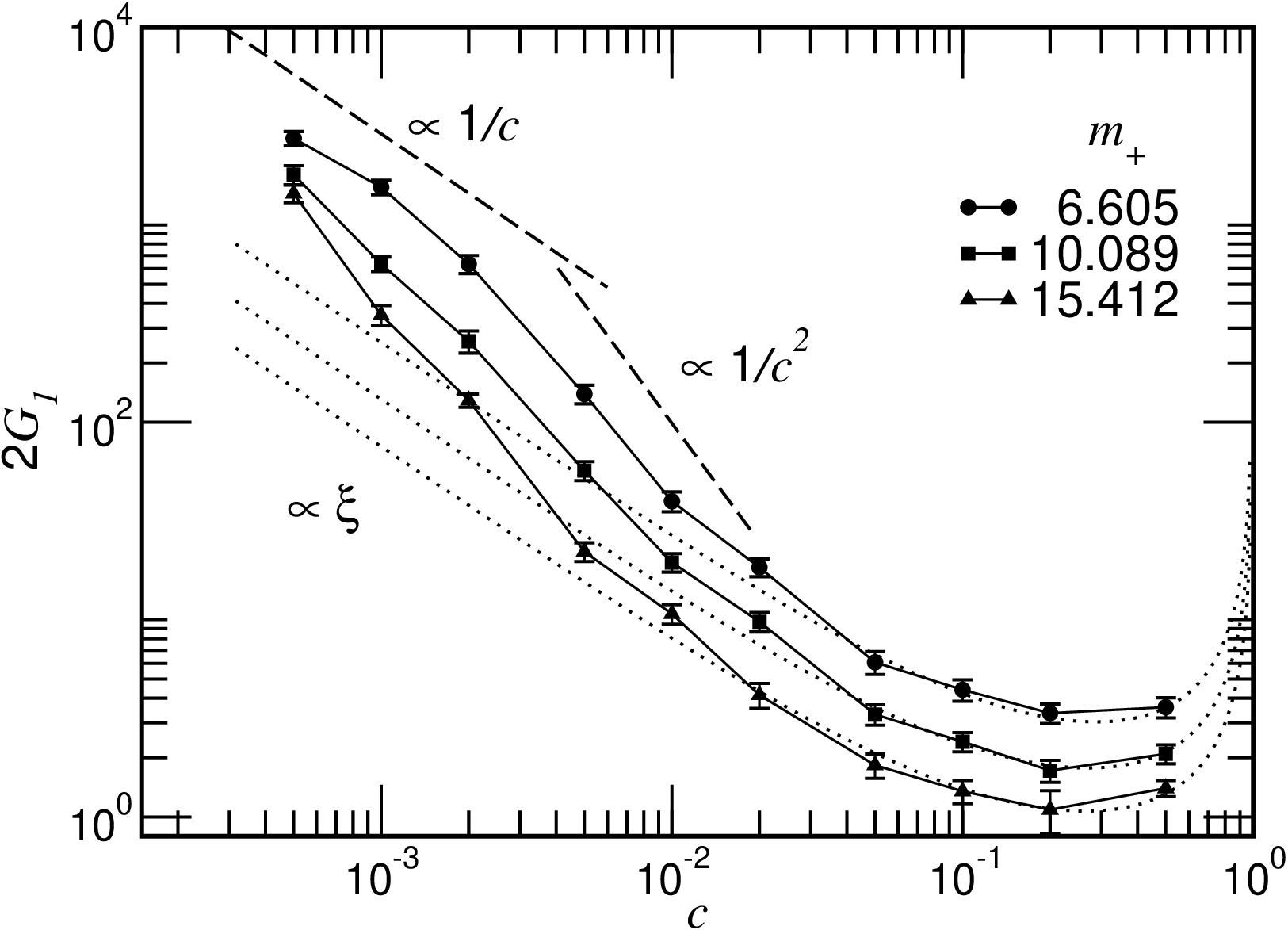}}{%
   \includegraphics[width=\columnwidth]{GRAPHS/TCoeff}}
  \caption{Transport coefficient $\TCoeff$ as a function of impurity 
	concentration $\Concentration$ for impurity cross 
	sections $\CrossSection =$ 0.30, 0.50, 0.70.
	Dashed line is proportional to $\Concentration^{-1}$.
	Dotted line is proportional to $\LocLength$.
  \FigSource{TCoeff}}
  \label{TCoeff:fig}
\end{figure}

Estimates of $2\TCoeff_1$ for all the systems considered are plotted in 
Fig.~\ref{TCoeff:fig} as a function of the impurity concentration 
$\Concentration$.  The calculated values appear as filled symbols having 
error bars that represent the SDM.  The thin
solid lines connecting the symbols are only to guide the eye.
The two dashed line segments appearing above the data indicate slopes 
that are proportional to $1/\Concentration$ and $1/\Concentration^2$.
The three dotted lines are proportional to $\LocLength$, as they appear in 
Fig.~\ref{xi_318:fig}.

From the data in Fig.~\ref{TCoeff:fig}, there appear to be three 
regions of interest.  At low concentration, the transport coefficient 
$\TCoeff_1$ is proportional to $\Concentration^{-1}$, as was expected from the 
discussion of time and length scales.  
As the impurity concentration increased, 
and the diffusion time 
$t_\LocLength$ decreases sufficiently that 
Eq.~(\ref{Dxi:eqn}) dominates, and $\TCoeff_1$ becomes proportional 
to $\Concentration^{-2}$.  
This transition is not apparent in the 
results of Payton \emph{et al.} \cite{Payton67} on a similar system 
having thermostats because their computing resourses prevented them 
from resolving the smaller
concentrations required to see the effect.

There is a noticeable difference in the rate at which 
the three curves make the transition from $\Concentration^{-1}$ to 
$\Concentration^{-2}$.  Based on the transition data for similar 
FPU-$\beta$ systems,
the transition rate should be inversely 
proportional to the impurity mass $\ImpurityMass$.
\cite{Snyder99}
Therefore, the transition rate for the $\MassPlus=15.412$ is slowing 
more rapidly, for a given change in impurity concentration, than the 
other two impuriy masses.   
As a result, the $\MassPlus=15.412$ systems do not exhibit distinct 
$1/\Concentration$ dependence at the concentrations studied.

The interesting behavior was the final transition at higher concentrations to 
a transport coefficient that is proportional to the initial localization 
length $\LocLength$. 
The three dotted curves labelled $\propto\LocLength$ are proportional to 
the localization lengths
that appear in Fig.~\ref{xi_318:fig} for $\Wavelength=31.8$.
The value of $\LocLength$
is multiplied by the same coefficient (approximately 0.09) to 
make the curves best agree with the measured transport coefficient.

\subsection{Localization Parameter}

The localization parameter can also be used as a measure of energy 
propagation.  Although systems of different lengths were used, 
energy transport, starting from a localized state, should be independent 
of total system length. 
Based on the previous discussion of the localization parameter,
the ratio $\Length/\LocParam$ is proportional to the 
number of masses over which the total energy is distributed.
For a given $\ImpurityMass$ and $\Concentration$, and assuming that 
$\Length\gg\LocLength$,
the number of masses, and not the fraction of the system 
($\LocParam_\infty/\LocParam$), 
should also be independent of total system length.
Therefore, the ratio $\Length/\LocParam$ will serve as a means for 
comparing results from systems of different lengths.

\begin{figure}
\ifthenelse{\boolean{Submittal}}{%
   \includegraphics[width=\columnwidth]{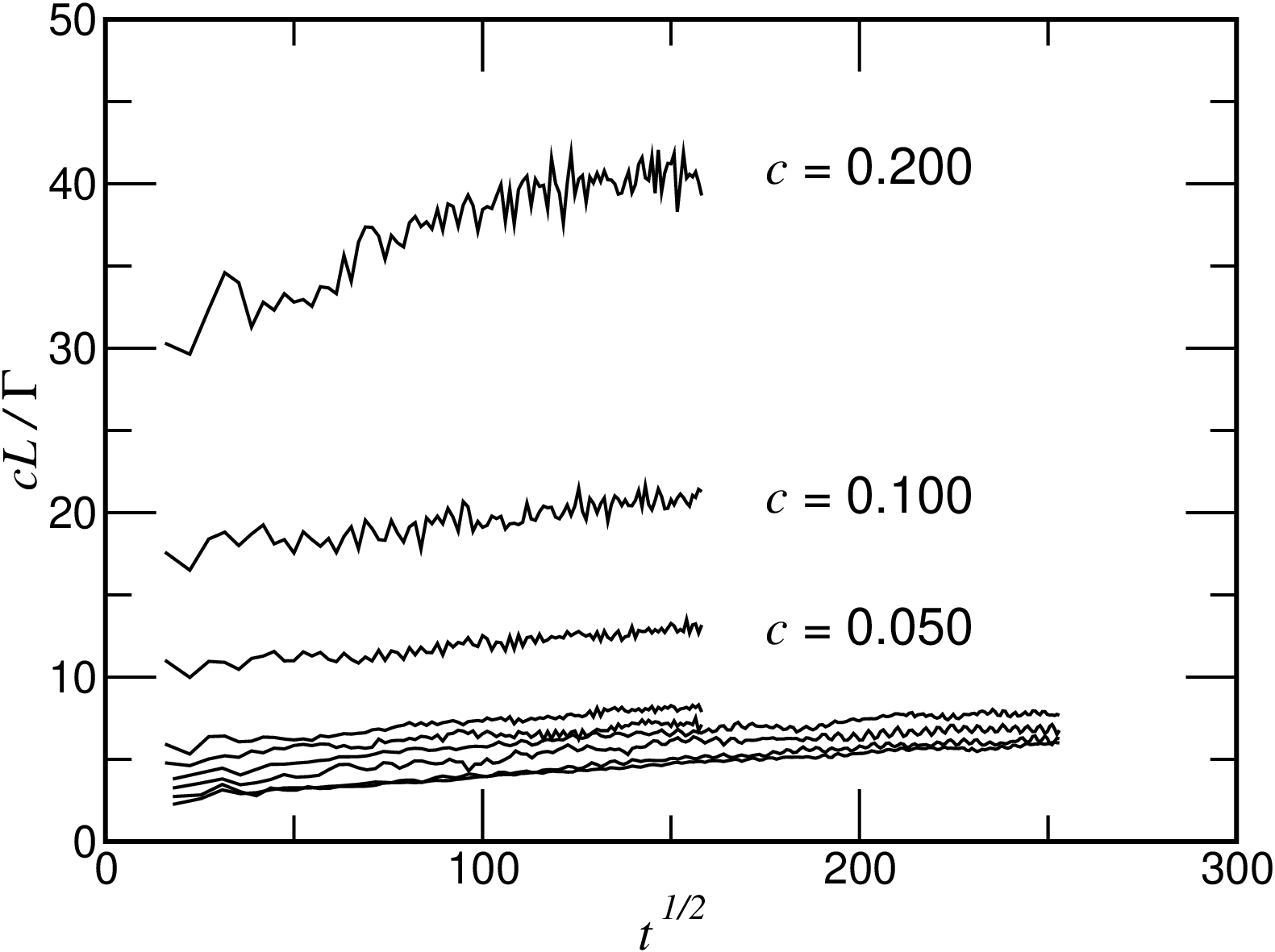}}{%
   \includegraphics[width=\columnwidth]{GRAPHS/LocParam30}}
  \caption{Localization parameter $\LocParam$ as a function of 
	time $t^{1/2}$ for systems having $\MassPlus =$ 6.605.
  \FigSource{LocParam30}}
  \label{LocParam30:fig}
\end{figure}

\begin{figure}
\ifthenelse{\boolean{Submittal}}{%
   \includegraphics[width=\columnwidth]{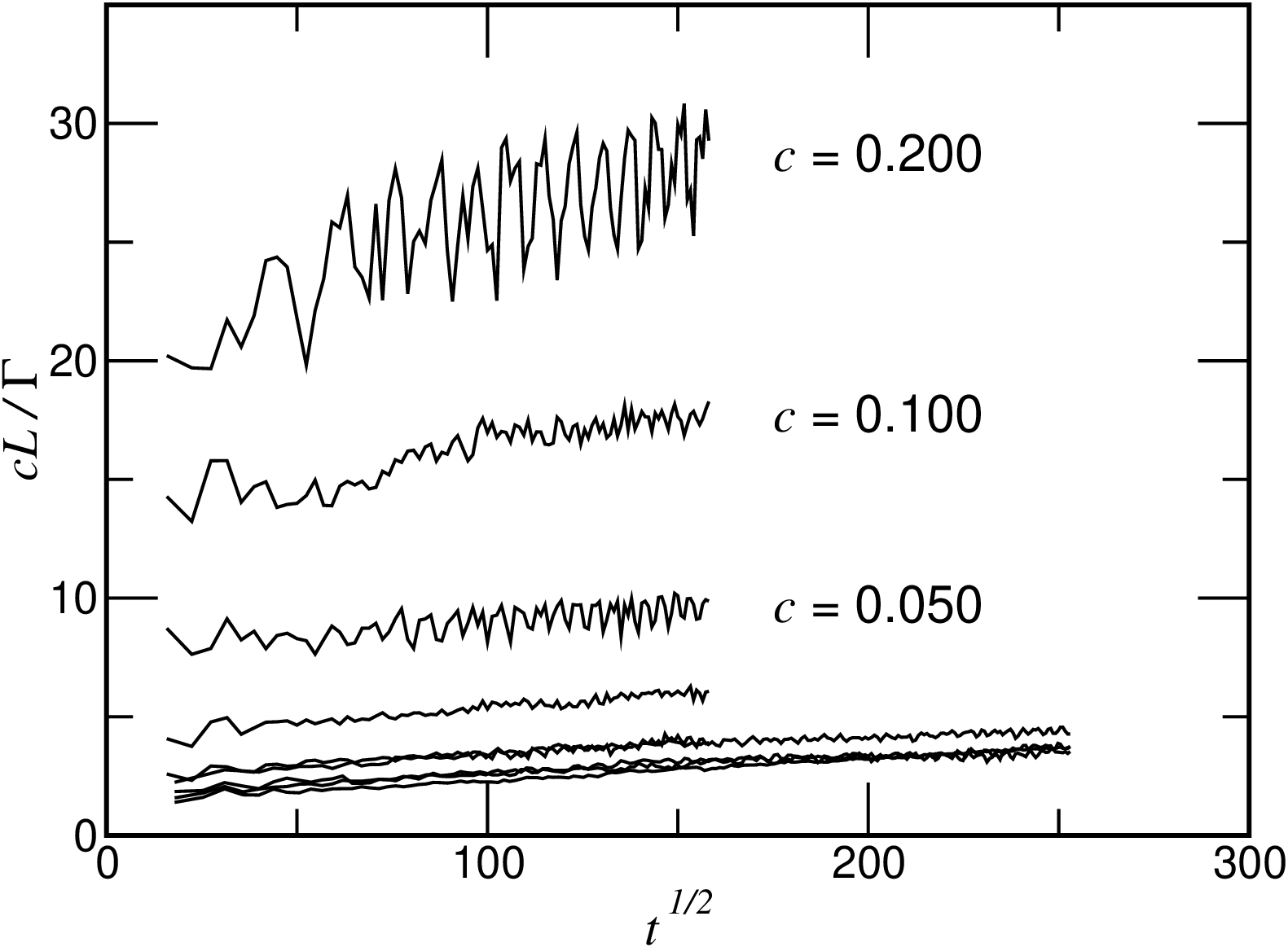}}{%
   \includegraphics[width=\columnwidth]{GRAPHS/LocParam50}}
  \caption{Localization parameter $\LocParam$ as a function of 
	time $t^{1/2}$ for systems having $\MassPlus =$ 10.089.
  \FigSource{LocParam50}}
  \label{LocParam50:fig}
\end{figure}

\begin{figure}
\ifthenelse{\boolean{Submittal}}{%
   \includegraphics[width=\columnwidth]{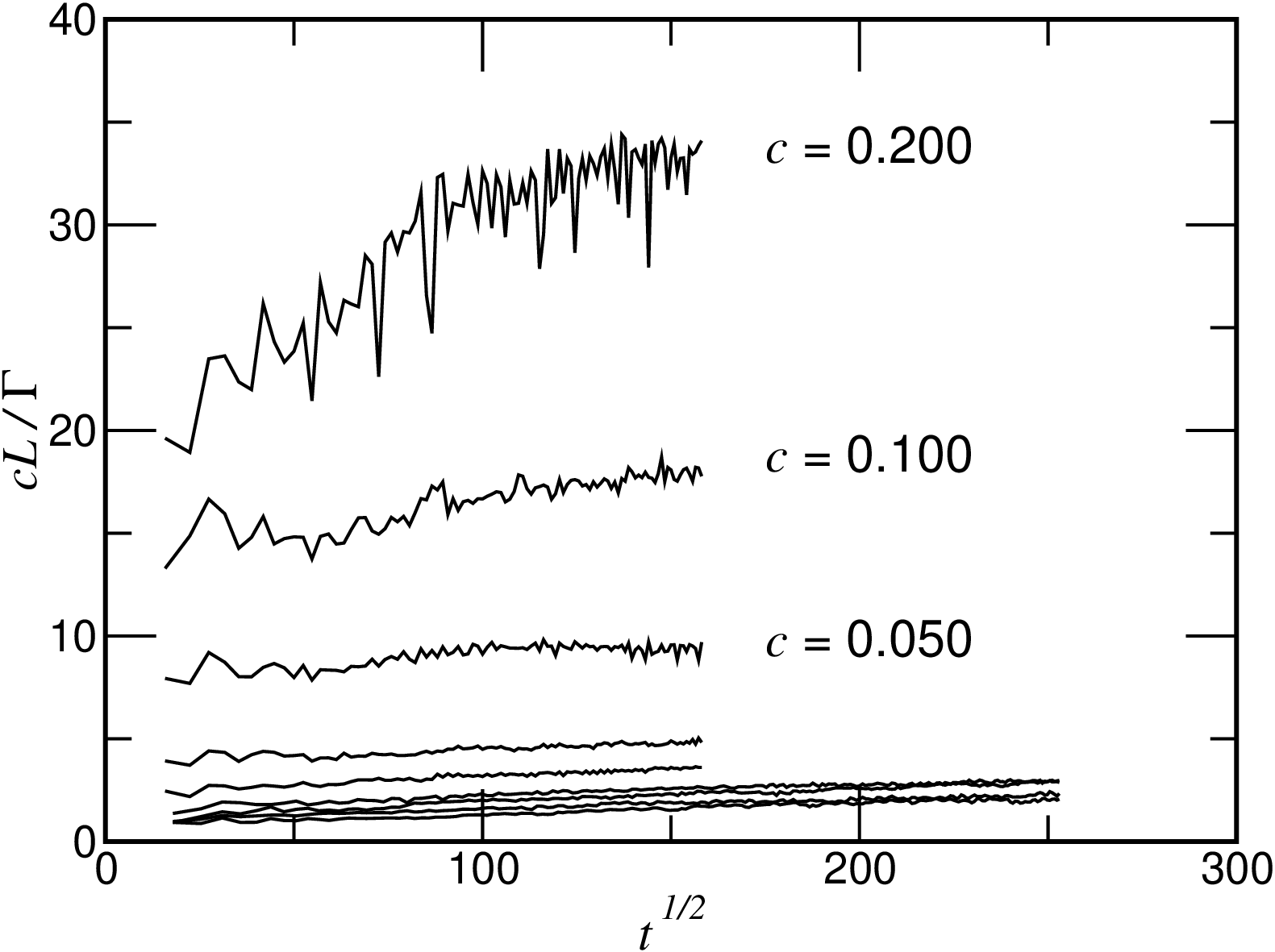}}{%
   \includegraphics[width=\columnwidth]{GRAPHS/LocParam70}}
  \caption{Localization parameter $\LocParam$ as a function of 
	time $t^{1/2}$ for systems having $\MassPlus =$ 15.412.
  \FigSource{LocParam70}}
  \label{LocParam70:fig}
\end{figure}

If the linear time dependence of the second moment $\Rmoment_1$ 
indicates diffusive behavior, the number of masses involved 
in energy transport should increase in proportion to $t^{1/2}$.
In addition, because transport in dilute systems is proportional 
to $\Concentration^{-1}$, the quantity $\Concentration\Length\LocParam$
should approach a constant for systems with low impurity concentrations.
The quantity $\Concentration\Length/\LocParam$, as a function of 
$t^{1/2}$, is plotted in Figs.~\ref{LocParam30:fig}--\ref{LocParam70:fig} 
for most of the systems studied.  In the figures, curves for smaller impurity 
concentration appear consecutively lower in the graph.  

There are two noteworthy features in 
Figs.~\ref{LocParam30:fig}--\ref{LocParam70:fig}.
The number of masses participating 
appears to be linear over the time intervals considered, particularly 
for the dilute impurity systems.
This is consistent
with the expectation of diffusive energy transport.
Also, as expected, the curves for the most dilute 
impurity concentrations appear to approach an asymptote, supporting 
the $\Concentration^{-1}$ dependence for the number of masses participating 
in energy transport.


\section{Discussion}

\subsection{$\beta$-Dependence}

\begin{figure}
\ifthenelse{\boolean{Submittal}}{%
   \includegraphics[width=\columnwidth]{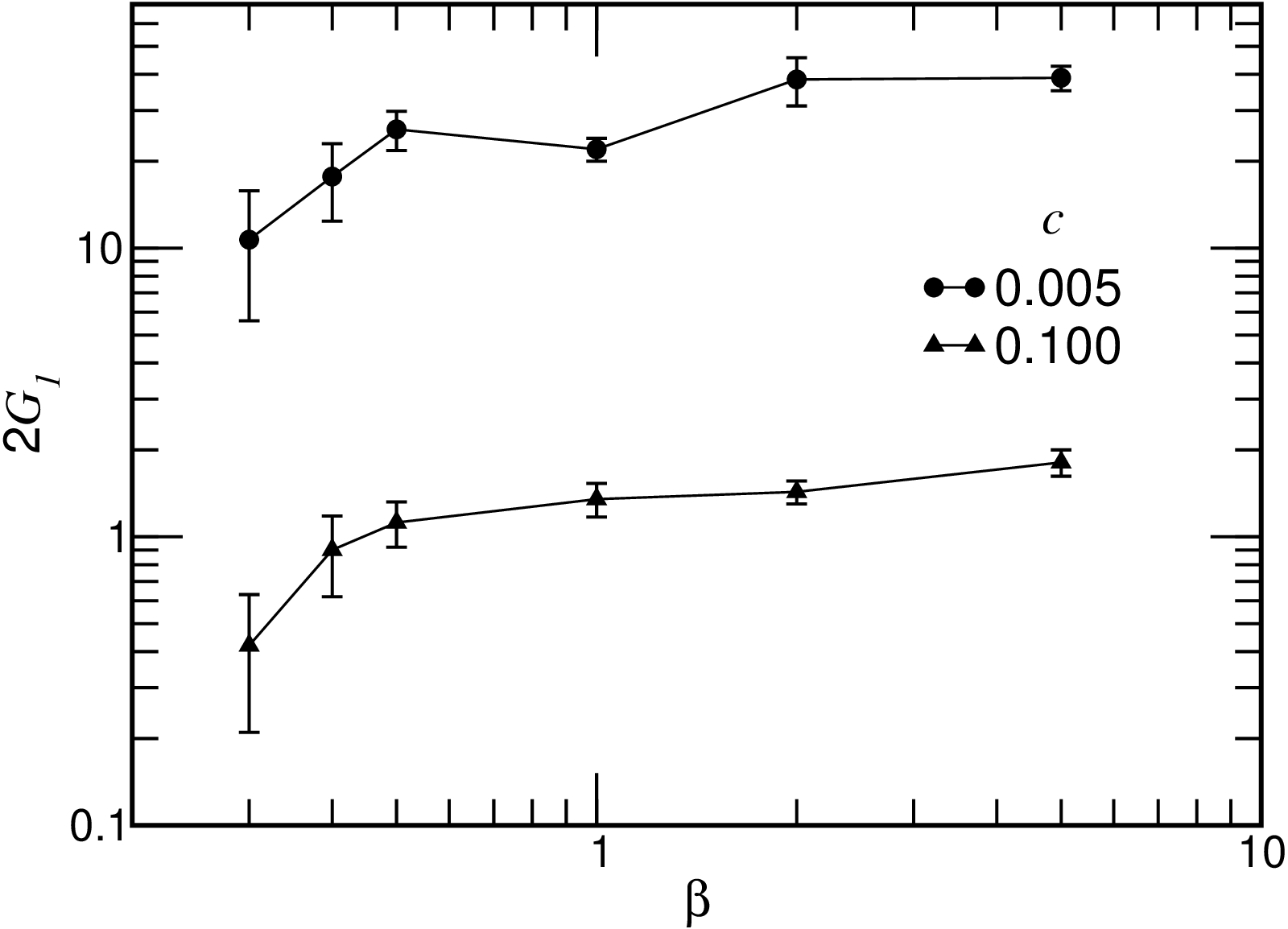}}{%
   \includegraphics[width=\columnwidth]{GRAPHS/Beta}}
  \caption{Transport coefficient $\TCoeff$ as a function of $\beta$ 
	for two impurity concentrations: $\Concentration =$ 0.100, 0.005,
	and $\CrossSection =$ 0.70.
   \FigSource{Beta}}
  \label{Beta:fig}	
\end{figure}

It was argued that the results given for $\beta$=1 are inicative of 
results for `large' values of $\beta$ that are above the critical 
threshold that leads to ergodic behavior.   As a check, the value of 
$\TCoeff_1$ for two systems are calculated for different values of 
$\beta$.  The two systems are 
$(\Concentration=0.005, \CrossSection=0.70)$ 
and 
$(\Concentration=0.100, \CrossSection=0.70)$.
The analysis for these systems was carried out in a manner identical 
to that for the $\beta=1$ data.  

The values for $\TCoeff_1$ are shown in Fig.~\ref{Beta:fig} as a 
function of $\beta$.  
Over the time interval studied, values of $\TCoeff_1$ decreased 
for values of $\beta$ below 0.5, 
falling to near zero at $\beta = 0.2$.
For $0.5\le\beta\le 5.0$, 
the transport coefficient has a weak dependence on the anharmonicity,
which is consistent with numerical experiments on finite temperature 
nonlinear BID systems.\cite{Payton67}
Therefore, while the calculated values of $\TCoeff$ are 
not constant, there does not appear to be any significance to 
a particular value for $\beta$ that is greater than 
0.5.

\subsection{Diffusion Time $t_\LocLength$}

The coefficient 
$2\TCoeff_1$ in 
Fig.~\ref{TCoeff:fig} represents the transport coefficient that 
characterizes the time $t_\LocLength$, defined in Eq.~(\ref{Dxi:eqn}):
\begin{equation}
	t_\LocLength \approx \LocLength^2 / \TCoeff_1
\end{equation}
At low concentration, because both $\TCoeff_1$ and $\LocLength$ are 
proportional to $\Concentration^{-1}$, the 
diffusion time $t_\LocLength$ must also be proportional to 
$\Concentration^{-1}$.  At intermediate concentration, $\TCoeff_1$ is 
proportional to $\Concentration^{-2}$, so $t_\LocLength$ is a weak 
function of impurity concentration.  At the highest concentrations 
studied, $\TCoeff_1$ and $t_\LocLength$ are both proportional 
to $\LocLength$.  At these concentrations, however, $\LocLength$ 
is a weak function of impurity concentration, so the same must 
be true for $t_\LocLength$.


\subsection{Time Scale}

\begin{figure}
\ifthenelse{\boolean{Submittal}}{%
   \includegraphics[width=\columnwidth]{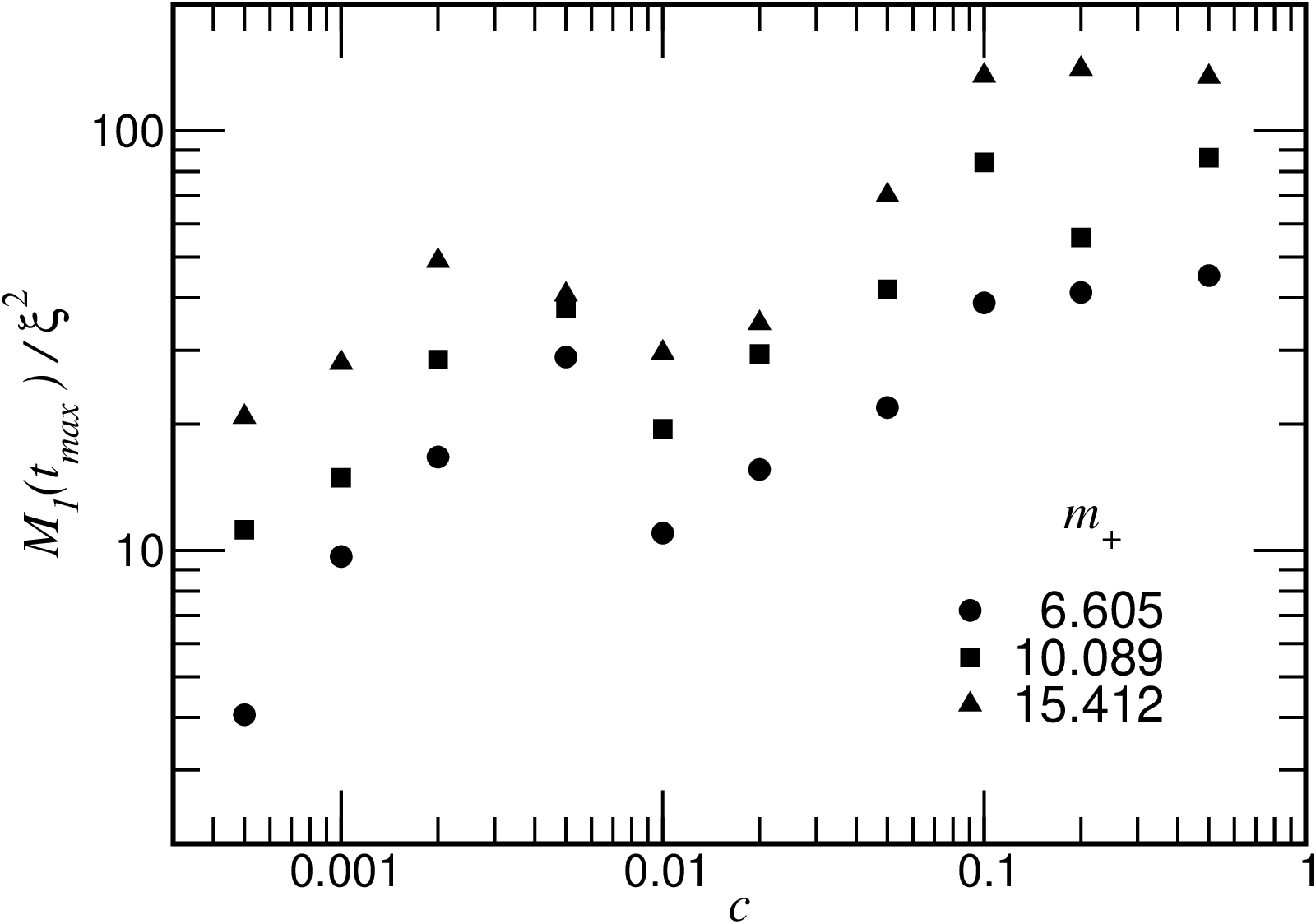}}{%
   \includegraphics[width=\columnwidth]{GRAPHS/M2MaxXi2}}
   \caption{The ratio $\Rmoment_1(t_{max})/\LocLength^2$ as a function 
	of impurity concentration $\Concentration$.
   \FigSource{M2MaxXi2}}
   \label{M2MaxXi2:fig}
\end{figure}

The significance of the reported results presented depends, in part, on the 
relative duration of the calculation.  One measure of duration, to gauge 
whether long times have been probed, is the ratio of the distance energy 
propagates along the system to its initial span.
The maximum energy propagation length is proportional to 
$\Rmoment_1(t_{max})$, and the initial span is proportional to the 
initial localization length $\LocLength$.

The ratio $\Rmoment_1(t_{max})/\LocLength^2$ 
for each system is plotted in Fig.~\ref{M2MaxXi2:fig}.  The ratio 
varied from 10 to 100, increasing with impurity concentration.
The square root of this ratio 
is a measure of the  depth to which energy propagated to the initial depth 
energy was distributed.  
Therefore, energy penetration depth, relative the initial localization 
length, varied from 3 to 10.  

Extending this experiment to longer times will probably yield similar 
results.  It is unlikely that ballistic transport will occur at long times.
As the calculation progresses, the energy is being distributed over 
a greater number of masses.  Because both energy and transition rates 
are proportional to the square of the oscillation amplitude, the mode 
transition rate is deceasing in proportion to $t^{-1/2}$

\subsection{Memory Effect}

Although previous arguments address why the transport coefficient 
should vary from $\Concentration^{-1}$ to $\Concentration^{-2}$,
they do not explain why the transport 
coefficient should be proportional to $\LocLength$ at the highest 
concentrations.  More specifically, it was unexpected that the 
transport coefficient is proportional to the original localization 
length and not some averaged value.  Figure~\ref{M2MaxXi2:fig} shows 
that the energy has propagated nearly ten times the localization length 
for the highest impurity concentrations, yet these systems seem to 
retains some ``memory'' of the initial state.

An explanation may be made from the results of the hoop experiment 
shown in Fig.~\ref{erg636_Neff_50:fig}.  For those systems, both the 
number of modes and the transition rate decreased significantly with 
increasing impurity concentration.  At these concentrations and time 
scales, mode generation may have been dominated by impurity 
scattering.
As the localization length is quite small for most modes, 
any new mode could 
only propagate a short distance before becoming localized.
If the impurities constrained mode generation sufficiently so that 
generated modes had frequencies near the initial mode, the 
initial localization length would remain as the controlling 
length scale over which new modes could propagate before scattering
again.

%
%

\section{Conclusion}

For systems initially localized near one end of a FPU-$\beta$ chain, 
anharmonicity facilitates energy transport along the chain.
For energy eigenstates of harmonic disordered chains, sufficient 
anharmonicity leads to diffusive energy transport.
The second moment of the site energies was linear over times long enough for
energy to have 
propagated a distance nearly equal to ten times the initial 
spatial extent of the pulse.
Although direct mass displacement observation showed low frequency 
modes propagating ballisticall, the energy content in these modes 
was insufficient for super-diffusive energy transport.

Alternate measures of energy and frequency distribution 
corroborated the assertion of diffusive behavior.
We found that the localization parameter was proportional to 
the square root of time, suggesting that energy 
distributes itself among masses in a diffusive manner.
This diffusive short range behavior is consistent with 
the observation of more long range diffusive behavior.
Calculations of the relative number of modes participating in 
periodic (hoop) systems suggest that an increase in the 
number of impurities reduces the number of active modes.

The most interesting aspect of the transport coefficient was the 
concentration dependence.
At the lowest impurity concentration $\Concentration$ values, 
the transport coefficient was proportional to $\Concentration^{-1}$.
At higher impurity concentrations, the transport coefficient 
developed a $\Concentration^{-2}$ dependence.
These two concentration dependences are consistent with 
arguments based on the competition between diffusion 
times and mode transition rates.

At the highest impurity concentrations, the transport 
coefficient was proportional to the original localization 
length.
The postulated cause for this was from the severe suppression of 
new frequency modes 
that occurs in concentrated BID systems.
These new modes, having a frequency in proximity of the 
original frequency, experienced a similar localization length, 
which dominated the distance travelled before subsequent 
scattering events.
As a result, the original localization length remained the dominant 
length scale over which transport occured for any given phonon.

\begin{acknowledgments}
The authors would like to thank Dr.\ Jack Douglas of the 
Polymers Division (NIST) for his 
useful comments and discussion.
This work was supported by the NSF under grant number 
DMR-01-32726, and by the National Institute of 
Standards and Technology (NIST) High Performance 
Construction Materials and Systems program 
in the Building and Fire Research Laboratory.
\end{acknowledgments}


\appendix*	

\section{Data Analysis}

The analyses of the results involve some minor subtleties that 
deserve a clear exposition.  Neglecting these subtleties and 
performing an ordinary least squares (OLS) analysis of the 
data would lead to misleading results.
Specifically, not adjusting for initial transient behavior would 
lead to a different conclusion regarding the existence of diffusive 
energy transport.  

As mentioned previously, the reported uncertainties are 
the standard deviation in the mean (SDM) calculated from the
ensemble population standard deviation $s$.  For 
an ensemble of $\Nens$ systems, the SDM reported here is 
$s/\sqrt{\Nens}$.  This uncertainty characterizes the 
reported average value from the population.

\subsection{Energy Fluctuation}

For each calculation, the total energy $\Etot(t)$ was calculated 
as a function of time.  Due to randomness, the initial total energy 
fluctuates among the ensembles.  To ensure that values of $\Etot$ 
were on comparable scales, the values were divided by the initial 
value $\Etot(0)$.   The ratio $\Etot_m(t)/\Etot_m(0)$ ($1\leq m\leq\Nens$) i
is calculated as a function of time 
for each of the $\Nens$ systems composing the ensemble.
The averaged values are calculated at each of the $P$ values of $t_i$:
\begin{equation}
	\overline{\Etot}(t_i) = \frac{1}{\Nens}\ \sum_{m}^{\Nens}
		\frac{\Etot_m(t_i)}{\Etot_m(0)}
	\hspace{2.0cm}
	1 \leq i\leq P
\end{equation}
These averaged values, along with the population standard deviation, 
are pooled and stored in the output data file.
The population of $\Nens$ values at $t_i$ do not, 
unfortunately, represent energy 
fluctuation for a single system.  Rather, it characterizes statistical 
fluctations among the different systems, evaluated at the same time.

Energy fluctuation can only be approximated from flucuations in the 
$P$ average values of $\overline{\Etot}(t_i)$.  
The standard deviation $s_{\overline{\Etot}}$ of 
each $\overline{\Etot}(t_i)$ represents a 
standard deviation in the mean.  The population standard deviation is 
approximated by $s_{\overline{\Etot}}\sqrt{P}$.
This population standard deviation was never more than 0.2~\%.  
Spot checks of $\Etot(t)$ in individual systems rarely 
gave a standard 
deviation greater than 0.2~\%.

%



\subsection{$\Texp_n$ Analysis}

\begin{figure}
\ifthenelse{\boolean{Submittal}}{%
   \includegraphics[width=\columnwidth]{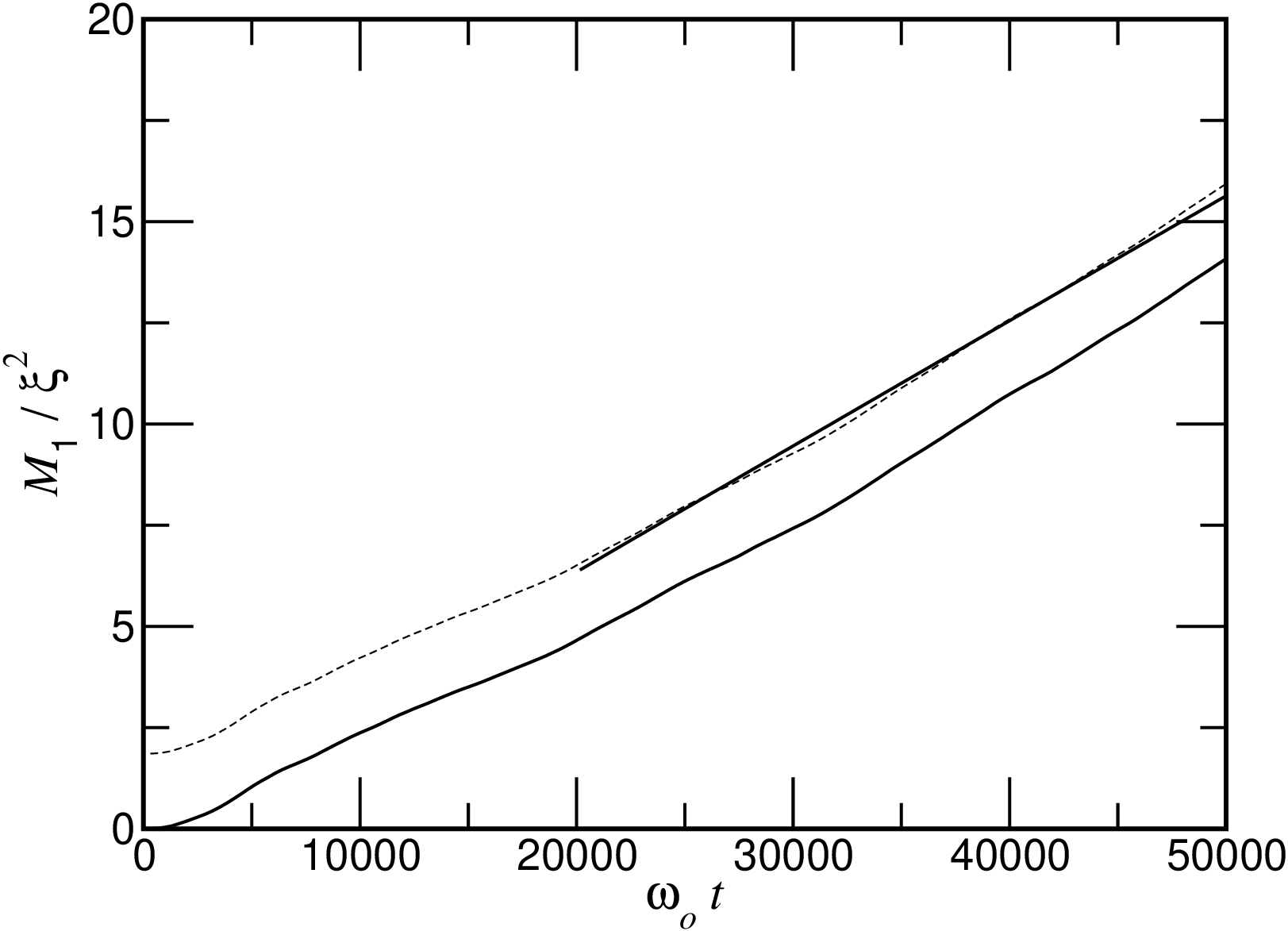}}{%
   \includegraphics[width=\columnwidth]{GRAPHS/diffusion}}
   \caption{$\Rmoment_1/\LocLength^2$ as a function of time.  
 	System length is 64~000, 
 	impurity concentration is 0.002, 
 	single impurity cross section is 0.30.
	The measured data appear as a solid line, 
	the shifted data appear as dashed line,
	and the line segment denotes the range over which regression 
	was performed.
   \FigSource{diffusion}}
   \label{diffusion:fig}
\end{figure}

The straightforward means of determining the time exponent 
$\Texp_n$ is from the slope of a log-log plot of $\Rmoment_n$ 
versus time $t$.  This approach assumes that the power law 
dependence exhibits itself at $t=0$.  In reality, there is a 
transient period, after which the value of $\Rmoment_n$ begins 
to increase.  Although this increase appears to be linear, 
the data must be corrected to eliminate the effect of the transient.

The effect of the transient behavior is negated by 
shifting the data for 
$\Rmoment_n$ vertically.  
OLS linear regression is applied to the $\Rmoment_n$ 
versus $t$
data over the time intervals given in Table~\ref{texp:table}.  The intercept 
calculated from the OLS regression is subtracted from 
$\Rmoment_n(t)$.  The process is demonstrated in Fig.~\ref{diffusion:fig}
for one particular system.  The original $\Rmoment_1$ data are shown as a solid 
line.  Regression is applied to the interval $[20000,50000]$, and the 
data are shifted vertically so that the linear approximation to this interval
has zero intercept.

OLS linear regression is then applied to the 
logarithm of these adjusted data versus the logarithm of time.
The slope of this regression calculation is the value 
reported in Figs.~\ref{Texp:fig} and \ref{TexpM2:fig} for 
$\Texp_1$ and $\Texp_2$, respectively.

\begin{figure}
\ifthenelse{\boolean{Submittal}}{%
   \includegraphics[width=\columnwidth]{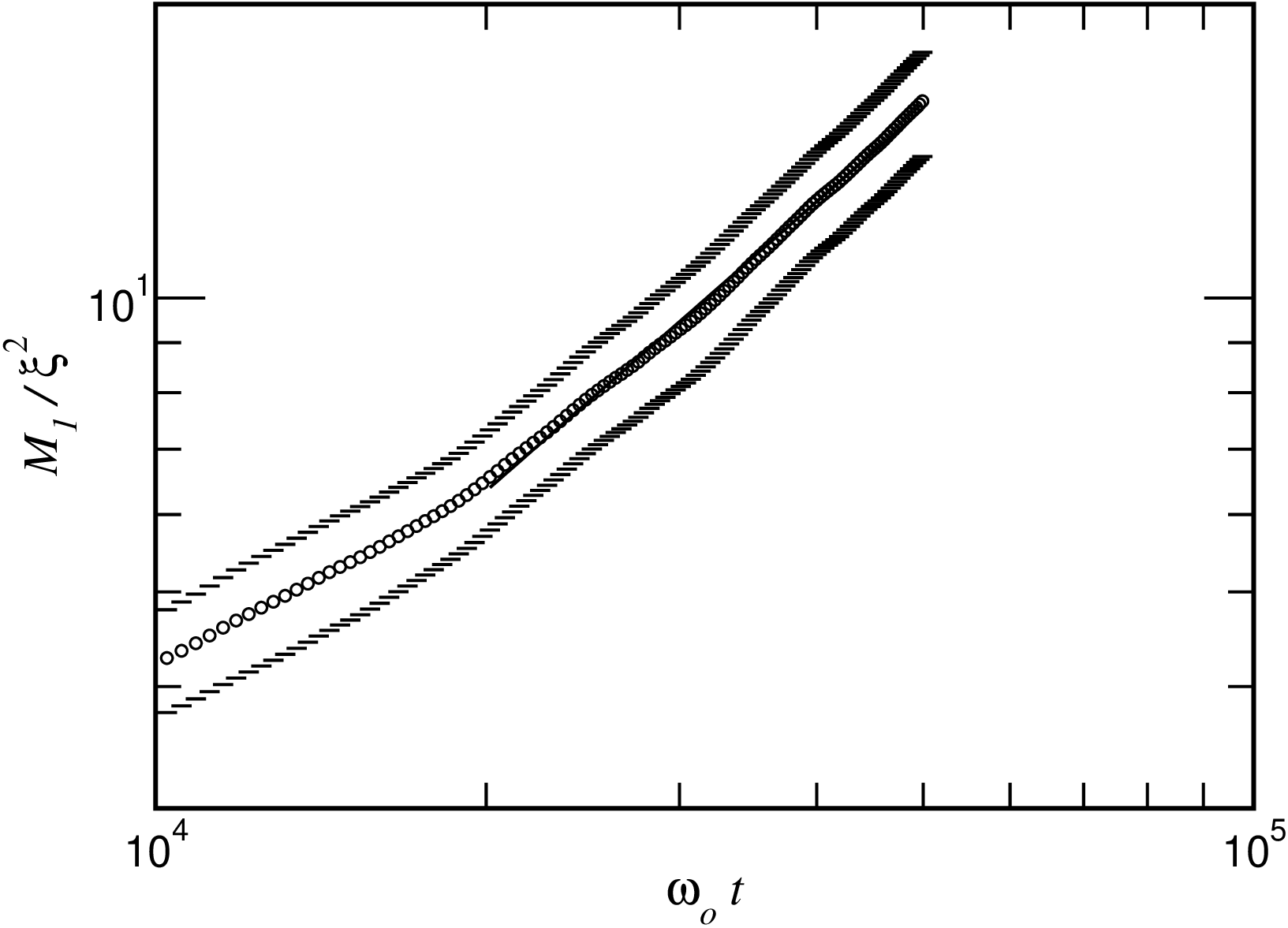}}{%
   \includegraphics[width=\columnwidth]{GRAPHS/slope}}
  \caption{A log-log plot of data in Fig.~\protect{\ref{diffusion:fig}}, 
	along with horizontal error bars denoting the population 
	standard deviation; the vertical risers are omitted for clarity.
	The solid line was determined by OLS regression applied to the mean 
	values.
  \FigSource{slope}}
  \label{slope:fig}
\end{figure}

The uncertainty in $\Texp_n$ is calculated from both the regression 
residuals and the ensemble population of $\Rmoment_n$ values.
Fortunately, because the population standard deviation in 
$\Rmoment_n$ increases in time, the logarithm transform yields 
uncertainties that are nearly constant over the time intervals of interest.
The adjusted data from Fig.~\ref{diffusion:fig} are shown in 
Fig.~\ref{slope:fig}, along with the ensemble population standard deviations 
that are denoted by horizizontal error bars; the vertical risers are 
omitted for clarity.  (The ensemble population standard deviation, 
instead of the SDM, is shown in the figure for clarity of the demonstration.
For an ensemble of $\Nens$ systems, the SDM error bars would be a factor 
of $\sqrt{\Nens}$ smaller than those shown in the figure.)
Also shown in the figure is the result of the regression analysis of the 
average values.  The error bars denote the ensemble uncertainty, and 
the residuals represent the regression uncertainty.

The OLS linear regression in log-space will yield an estimated 
uncertainty (standard deviation) for slope that is a function of the 
standard error of the residuals.  The standard error $s$ 
can be used to estimate the 
regression standard deviation $s_{reg}$ for the slope $\Texp_n$:  
\begin{equation}
	s_{reg} = \frac{s}{\sqrt{S_{XX}}}
\end{equation}
where the quantity $S_{XX}$ is the sum of squares:
\begin{equation}
	S_{XX} = \sum_i \ \left(\ln t_i - \overline{\ln t}\right)^2
\end{equation}
The quantity $\overline{\ln t}$ represents the average value over the 
specified interval.

The uncertainty in $\Texp_n$ should also reflect the SDM recorded for the 
ensemble $s_{ens}$. 
It is assumed that these two uncertainties 
are independent of one another, and that they are additive.
The uncertainty (estimated standard deviation) in the time 
exponent is 
\begin{equation}
	s_{\Texp_n}^2  = \frac{s_{res}^2}{S_{XX}} + \frac{s_{ens}^2}{S_{XX}}
\end{equation}
Because $s_{ens}$ is the majority of $s_{\Texp_n}$, the uncertainty 
in $\Texp_n$ is referred to as a SDM, and the 
coverage factor is approximately equivalent to one standard deviation 
for a normal distribution.


\subsection{$\Rmoment_1$ Analysis}

Based on the results of the $\Texp_1$ analysis, a hypothesis of a 
linear relationship between the second moment $\Rmoment_1$ and time $t$ 
cannot be rejected.  The ability to use a linear model simplifies the 
analysis of $\Rmoment_1$.  The transport coefficient $2\TCoeff_1$ is the 
slope that is calculated from
OLS regression using a linear model.

The uncertainty in $2\TCoeff_1$ cannot be determined from a regression 
analysis of the $\Rmoment_1$ versus $t$ data because the uncertainties 
in $\Rmoment_1$ increase in time.
Fortunately, the data can be transformed into a more suitable format.  

Assuming diffusive behavior, the values of $\Rmoment_1$ are a collection of 
lines, radiating from the origin.  
The error model for the observations assumes that there exists an 
inherent error $\Error$ and that the total error increases with 
time:
\begin{equation}
	\Rmoment_n (t_i) = A + 2\TCoeff_n\, t_i + t_i \Error_i
	\label{error1:eqn}
\end{equation}
To use OLS techniques, the error term must be additive and 
constant.
Equation~(\ref{error1:eqn}) can be transformed into a suitable model:
\begin{equation}
	\frac{\Rmoment_n (t_i)}{t_i}
		= \frac{A}{t_i} + 2\TCoeff_n + \Error_i
	\label{Gerror:eqn}
\end{equation}

Unfortunately, the initial transient behavior that confounded
the calculation of $\Texp_n$ must also be accounted for here.
The error model assumes that the uncertainty grows linearly 
from $t=0$.  
In reality, the increase in $\Rmoment_n$ occured after some initial transient
time $t_\circ$.
To correct for this, linear regression is applied to the 
$\Rmoment_1$ versus $t$ data to determine the value of 
$t_\circ$.  Using $t_\circ$, 
the data are shifted horizontally so that the linear region of 
interest points to the origin.  These
shifted data are then transformed according to Eq.~(\ref{Gerror:eqn}).

\begin{figure}
\ \vspace{0.5cm}
\ifthenelse{\boolean{Submittal}}{%
   \includegraphics[width=\columnwidth]{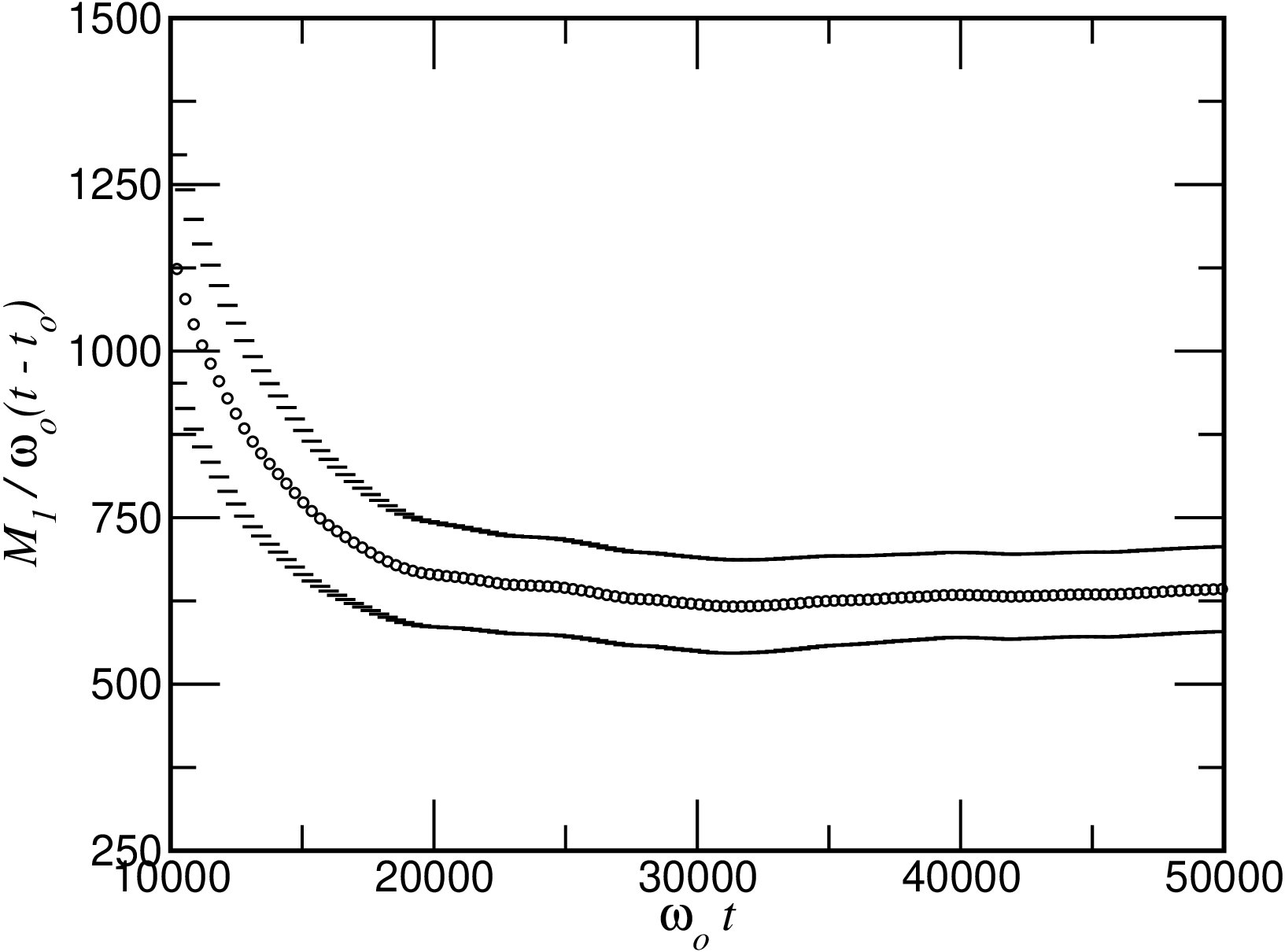}}{%
   \includegraphics[width=\columnwidth]{GRAPHS/Gcalc}}
  \caption{Transformed $\Rmoment_1$ to demonstrate how uncertainty 
	in $\Texp_1$ is calculated. 
  \FigSource{Gcalc}}
  \label{Gcalc:fig}
\end{figure}

The result of this transform, applied to both the mean and SDM, 
is shown in Fig.~\ref{Gcalc:fig} for the 
data in Figs.~\ref{diffusion:fig} and \ref{slope:fig}.  
In Fig.~\ref{Gcalc:fig}, the error bars represent the SDM.
Over the range of regression (see Table~\ref{texp:table}), the 
SDM is relatively constant, consistent with Eq.~(\ref{Gerror:eqn}).  
Because the model in Eqn.~(\ref{Gerror:eqn}) assumes a constant 
factor of $2\TCoeff_1$, 
the error bars in Fig.~\ref{Gcalc:fig} represent the SDM for 
$2\TCoeff_1$.

\bibliography{/Users/jackal/jdb/j_abbrev,references}

\begin{thebibliography}{39}
\expandafter\ifx\csname natexlab\endcsname\relax\def\natexlab#1{#1}\fi
\expandafter\ifx\csname bibnamefont\endcsname\relax
  \def\bibnamefont#1{#1}\fi
\expandafter\ifx\csname bibfnamefont\endcsname\relax
  \def\bibfnamefont#1{#1}\fi
\expandafter\ifx\csname citenamefont\endcsname\relax
  \def\citenamefont#1{#1}\fi
\expandafter\ifx\csname url\endcsname\relax
  \def\url#1{\texttt{#1}}\fi
\expandafter\ifx\csname urlprefix\endcsname\relax\def\urlprefix{URL }\fi
\providecommand{\bibinfo}[2]{#2}
\providecommand{\eprint}[2][]{\url{#2}}

\bibitem[{\citenamefont{Held et~al.}(1997)\citenamefont{Held, Pfeiffer, and
  Kuhn}}]{Held97}
\bibinfo{author}{\bibfnamefont{T.}~\bibnamefont{Held}},
  \bibinfo{author}{\bibfnamefont{I.}~\bibnamefont{Pfeiffer}}, \bibnamefont{and}
  \bibinfo{author}{\bibfnamefont{W.}~\bibnamefont{Kuhn}},
  \bibinfo{journal}{Phys.\ Rev.\ B} \textbf{\bibinfo{volume}{55}},
  \bibinfo{pages}{231} (\bibinfo{year}{1997}).

\bibitem[{\citenamefont{Rohmfeld et~al.}(2001)\citenamefont{Rohmfeld,
  Hundhausen, Ley, Schulze, and Pensl}}]{Rohmfeld01}
\bibinfo{author}{\bibfnamefont{S.}~\bibnamefont{Rohmfeld}},
  \bibinfo{author}{\bibfnamefont{M.}~\bibnamefont{Hundhausen}},
  \bibinfo{author}{\bibfnamefont{L.}~\bibnamefont{Ley}},
  \bibinfo{author}{\bibfnamefont{N.}~\bibnamefont{Schulze}}, \bibnamefont{and}
  \bibinfo{author}{\bibfnamefont{G.}~\bibnamefont{Pensl}},
  \bibinfo{journal}{Phys.\ Rev.\ Lett.} \textbf{\bibinfo{volume}{86}},
  \bibinfo{pages}{826} (\bibinfo{year}{2001}).

\bibitem[{\citenamefont{Widulle et~al.}(2002)\citenamefont{Widulle, Serrano,
  and Cardona}}]{Widulle02}
\bibinfo{author}{\bibfnamefont{F.}~\bibnamefont{Widulle}},
  \bibinfo{author}{\bibfnamefont{J.}~\bibnamefont{Serrano}}, \bibnamefont{and}
  \bibinfo{author}{\bibfnamefont{M.}~\bibnamefont{Cardona}},
  \bibinfo{journal}{Phys.\ Rev.\ B} \textbf{\bibinfo{volume}{65}},
  \bibinfo{pages}{075206} (\bibinfo{year}{2002}).

\bibitem[{\citenamefont{Wagner et~al.}(1992)\citenamefont{Wagner, Zavt,
  Vazquez-Marquez, L{\"u}tze, Mougios, Viliani, Frizzera, Pilla, and
  Montagna}}]{Wagner92}
\bibinfo{author}{\bibfnamefont{M.}~\bibnamefont{Wagner}},
  \bibinfo{author}{\bibfnamefont{G.}~\bibnamefont{Zavt}},
  \bibinfo{author}{\bibfnamefont{J.}~\bibnamefont{Vazquez-Marquez}},
  \bibinfo{author}{\bibfnamefont{A.}~\bibnamefont{L{\"u}tze}},
  \bibinfo{author}{\bibfnamefont{T.}~\bibnamefont{Mougios}},
  \bibinfo{author}{\bibfnamefont{G.}~\bibnamefont{Viliani}},
  \bibinfo{author}{\bibfnamefont{W.}~\bibnamefont{Frizzera}},
  \bibinfo{author}{\bibfnamefont{O.}~\bibnamefont{Pilla}}, \bibnamefont{and}
  \bibinfo{author}{\bibfnamefont{M.}~\bibnamefont{Montagna}},
  \bibinfo{journal}{Philos.\ Mag.\ B} \textbf{\bibinfo{volume}{65}},
  \bibinfo{pages}{273} (\bibinfo{year}{1992}).

\bibitem[{\citenamefont{Solodov and Korshak}(2002)}]{Solodov02}
\bibinfo{author}{\bibfnamefont{I.~Y.} \bibnamefont{Solodov}} \bibnamefont{and}
  \bibinfo{author}{\bibfnamefont{B.~A.} \bibnamefont{Korshak}},
  \bibinfo{journal}{Phys.\ Rev.\ Lett.} \textbf{\bibinfo{volume}{88}},
  \bibinfo{pages}{014303} (\bibinfo{year}{2002}).

\bibitem[{\citenamefont{Solodov et~al.}(2004)\citenamefont{Solodov, Wackerl,
  Pfleiderer, and Busse}}]{Solodov04}
\bibinfo{author}{\bibfnamefont{I.}~\bibnamefont{Solodov}},
  \bibinfo{author}{\bibfnamefont{J.}~\bibnamefont{Wackerl}},
  \bibinfo{author}{\bibfnamefont{K.}~\bibnamefont{Pfleiderer}},
  \bibnamefont{and} \bibinfo{author}{\bibfnamefont{G.}~\bibnamefont{Busse}},
  \bibinfo{journal}{Appl.\ Phys.\ Lett.} \textbf{\bibinfo{volume}{84}},
  \bibinfo{pages}{5386} (\bibinfo{year}{2004}).

\bibitem[{\citenamefont{Anderson}(1958)}]{Anderson58}
\bibinfo{author}{\bibfnamefont{P.~W.} \bibnamefont{Anderson}},
  \bibinfo{journal}{Phys.\ Rev.} \textbf{\bibinfo{volume}{109}},
  \bibinfo{pages}{1492} (\bibinfo{year}{1958}).

\bibitem[{\citenamefont{Bourbonnais and
  Maynard}(1990{\natexlab{a}})}]{Bourbonnais90a}
\bibinfo{author}{\bibfnamefont{R.}~\bibnamefont{Bourbonnais}} \bibnamefont{and}
  \bibinfo{author}{\bibfnamefont{R.}~\bibnamefont{Maynard}},
  \bibinfo{journal}{Phys.\ Rev.\ Lett.} \textbf{\bibinfo{volume}{64}},
  \bibinfo{pages}{1397} (\bibinfo{year}{1990}{\natexlab{a}}).

\bibitem[{\citenamefont{Bourbonnais and
  Maynard}(1990{\natexlab{b}})}]{Bourbonnais90b}
\bibinfo{author}{\bibfnamefont{R.}~\bibnamefont{Bourbonnais}} \bibnamefont{and}
  \bibinfo{author}{\bibfnamefont{R.}~\bibnamefont{Maynard}},
  \bibinfo{journal}{Int.\ J.\ Mod.\ Phys.\ C} \textbf{\bibinfo{volume}{1}},
  \bibinfo{pages}{233} (\bibinfo{year}{1990}{\natexlab{b}}).

\bibitem[{\citenamefont{Sarmiento et~al.}(1999)\citenamefont{Sarmiento,
  Reigada, Romero, and Lindenberg}}]{Sarmiento99}
\bibinfo{author}{\bibfnamefont{A.}~\bibnamefont{Sarmiento}},
  \bibinfo{author}{\bibfnamefont{R.}~\bibnamefont{Reigada}},
  \bibinfo{author}{\bibfnamefont{A.~H.} \bibnamefont{Romero}},
  \bibnamefont{and}
  \bibinfo{author}{\bibfnamefont{K.}~\bibnamefont{Lindenberg}},
  \bibinfo{journal}{Phys.\ Rev.\ E} \textbf{\bibinfo{volume}{60}},
  \bibinfo{pages}{5317} (\bibinfo{year}{1999}).

\bibitem[{\citenamefont{Yamada and Ikeda}(1999)}]{Yamada99}
\bibinfo{author}{\bibfnamefont{H.}~\bibnamefont{Yamada}} \bibnamefont{and}
  \bibinfo{author}{\bibfnamefont{K.~S.} \bibnamefont{Ikeda}},
  \bibinfo{journal}{Phys.\ Rev.\ E} \textbf{\bibinfo{volume}{59}},
  \bibinfo{pages}{5214} (\bibinfo{year}{1999}).

\bibitem[{\citenamefont{Rosas and Lindenberg}(2004)}]{Rosas04}
\bibinfo{author}{\bibfnamefont{A.}~\bibnamefont{Rosas}} \bibnamefont{and}
  \bibinfo{author}{\bibfnamefont{K.}~\bibnamefont{Lindenberg}},
  \bibinfo{journal}{Phys.\ Rev.\ E} \textbf{\bibinfo{volume}{69}}
  (\bibinfo{year}{2004}).

\bibitem[{\citenamefont{Fermi et~al.}(1955)\citenamefont{Fermi, Pasta, and
  Ulam}}]{Fermi55}
\bibinfo{author}{\bibfnamefont{E.}~\bibnamefont{Fermi}},
  \bibinfo{author}{\bibfnamefont{J.}~\bibnamefont{Pasta}}, \bibnamefont{and}
  \bibinfo{author}{\bibfnamefont{S.}~\bibnamefont{Ulam}}, \bibinfo{type}{Tech.
  Rep.} \bibinfo{number}{LA-1940}, \bibinfo{institution}{Los Alamos Scientific
  Laboratory} (\bibinfo{year}{1955}), \bibinfo{note}{{S}ee also: \emph{The
  Many-Body Problem}, edited by D.C. Mattis, World Scientific, 851--870,
  (1993)}.

\bibitem[{\citenamefont{Helfand}(1960)}]{Helfand60}
\bibinfo{author}{\bibfnamefont{E.}~\bibnamefont{Helfand}},
  \bibinfo{journal}{Phys. Rev.} \textbf{\bibinfo{volume}{119}},
  \bibinfo{pages}{1} (\bibinfo{year}{1960}).

\bibitem[{\citenamefont{Cretegny et~al.}(1998)\citenamefont{Cretegny, Dauxois,
  Ruffo, and Torcini}}]{Cretegny98}
\bibinfo{author}{\bibfnamefont{T.}~\bibnamefont{Cretegny}},
  \bibinfo{author}{\bibfnamefont{T.}~\bibnamefont{Dauxois}},
  \bibinfo{author}{\bibfnamefont{S.}~\bibnamefont{Ruffo}}, \bibnamefont{and}
  \bibinfo{author}{\bibfnamefont{A.}~\bibnamefont{Torcini}},
  \bibinfo{journal}{Physica D} \textbf{\bibinfo{volume}{121}},
  \bibinfo{pages}{106} (\bibinfo{year}{1998}), \eprint{cond-mat/9709204}.

\bibitem[{\citenamefont{Piazza et~al.}(2001)\citenamefont{Piazza, Lepri, and
  Livi}}]{Piazza01}
\bibinfo{author}{\bibfnamefont{F.}~\bibnamefont{Piazza}},
  \bibinfo{author}{\bibfnamefont{S.}~\bibnamefont{Lepri}}, \bibnamefont{and}
  \bibinfo{author}{\bibfnamefont{R.}~\bibnamefont{Livi}}, \bibinfo{journal}{J.\
  Phys.\ A} \textbf{\bibinfo{volume}{34}}, \bibinfo{pages}{9803}
  (\bibinfo{year}{2001}).

\bibitem[{\citenamefont{Yoshida}(1990)}]{Yoshida90}
\bibinfo{author}{\bibfnamefont{H.}~\bibnamefont{Yoshida}},
  \bibinfo{journal}{Phys.\ Lett.\ A} \textbf{\bibinfo{volume}{150}},
  \bibinfo{pages}{262} (\bibinfo{year}{1990}).

\bibitem[{\citenamefont{Livi et~al.}(1985)\citenamefont{Livi, Pettini, Ruffo,
  Sparpaglione, and Vulpiani}}]{Livi85}
\bibinfo{author}{\bibfnamefont{R.}~\bibnamefont{Livi}},
  \bibinfo{author}{\bibfnamefont{M.}~\bibnamefont{Pettini}},
  \bibinfo{author}{\bibfnamefont{S.}~\bibnamefont{Ruffo}},
  \bibinfo{author}{\bibfnamefont{M.}~\bibnamefont{Sparpaglione}},
  \bibnamefont{and} \bibinfo{author}{\bibfnamefont{A.}~\bibnamefont{Vulpiani}},
  \bibinfo{journal}{Phys.\ Rev.\ A} \textbf{\bibinfo{volume}{31}},
  \bibinfo{pages}{1039} (\bibinfo{year}{1985}).

\bibitem[{\citenamefont{Khomeriki et~al.}(2004)\citenamefont{Khomeriki, Lepri,
  and Ruffo}}]{Khomeriki04}
\bibinfo{author}{\bibfnamefont{R.}~\bibnamefont{Khomeriki}},
  \bibinfo{author}{\bibfnamefont{S.}~\bibnamefont{Lepri}}, \bibnamefont{and}
  \bibinfo{author}{\bibfnamefont{S.}~\bibnamefont{Ruffo}}
  (\bibinfo{year}{2004}), \eprint{cond-mat/0407134}.

\bibitem[{\citenamefont{de~L.~Kronig and Penney}(1931)}]{Kronig31}
\bibinfo{author}{\bibfnamefont{R.}~\bibnamefont{de~L.~Kronig}}
  \bibnamefont{and} \bibinfo{author}{\bibfnamefont{W.~G.}
  \bibnamefont{Penney}}, \bibinfo{journal}{Proc.\ Roy.\ Soc.\ Lond.\ A}
  \textbf{\bibinfo{volume}{130}}, \bibinfo{pages}{499} (\bibinfo{year}{1931}).

\bibitem[{\citenamefont{Ziman}(1979)}]{Ziman79}
\bibinfo{author}{\bibfnamefont{J.~M.} \bibnamefont{Ziman}},
  \emph{\bibinfo{title}{Models of Disorder}} (\bibinfo{publisher}{Cambridge
  University Press}, \bibinfo{address}{Cambridge}, \bibinfo{year}{1979}).

\bibitem[{\citenamefont{Morse and Ingard}(1968)}]{Morse68}
\bibinfo{author}{\bibfnamefont{P.~M.} \bibnamefont{Morse}} \bibnamefont{and}
  \bibinfo{author}{\bibfnamefont{K.~U.} \bibnamefont{Ingard}},
  \emph{\bibinfo{title}{Theoretical Acoustics}} (\bibinfo{publisher}{Princeton
  University Press}, \bibinfo{year}{1968}).

\bibitem[{\citenamefont{Snyder and Kirkpatrick}(2004)}]{Snyder04}
\bibinfo{author}{\bibfnamefont{K.~A.} \bibnamefont{Snyder}} \bibnamefont{and}
  \bibinfo{author}{\bibfnamefont{T.~R.} \bibnamefont{Kirkpatrick}},
  \bibinfo{journal}{Phys.\ Rev.\ B} \textbf{\bibinfo{volume}{70}},
  \bibinfo{pages}{104201} (\bibinfo{year}{2004}).

\bibitem[{\citenamefont{Kittel}(1986)}]{Kittel86}
\bibinfo{author}{\bibfnamefont{C.}~\bibnamefont{Kittel}},
  \emph{\bibinfo{title}{Introduction to Solid State Physics}}
  (\bibinfo{publisher}{Wiley}, \bibinfo{year}{1986}), \bibinfo{edition}{sixth}
  ed.

\bibitem[{\citenamefont{Anderson et~al.}(1980)\citenamefont{Anderson, Thouless,
  Abrahams, and Fisher}}]{Anderson80}
\bibinfo{author}{\bibfnamefont{P.~W.} \bibnamefont{Anderson}},
  \bibinfo{author}{\bibfnamefont{D.~J.} \bibnamefont{Thouless}},
  \bibinfo{author}{\bibfnamefont{E.}~\bibnamefont{Abrahams}}, \bibnamefont{and}
  \bibinfo{author}{\bibfnamefont{D.~S.} \bibnamefont{Fisher}},
  \bibinfo{journal}{Phys.\ Rev.\ B} \textbf{\bibinfo{volume}{22}},
  \bibinfo{pages}{3519} (\bibinfo{year}{1980}).

\bibitem[{\citenamefont{Landauer}(1970)}]{Landauer70}
\bibinfo{author}{\bibfnamefont{R.}~\bibnamefont{Landauer}},
  \bibinfo{journal}{Philos.\ Mag.} \textbf{\bibinfo{volume}{21}},
  \bibinfo{pages}{863} (\bibinfo{year}{1970}).

\bibitem[{\citenamefont{Payton et~al.}(1967)\citenamefont{Payton, Rich, and
  Visscher}}]{Payton67}
\bibinfo{author}{\bibfnamefont{D.~N.} \bibnamefont{Payton}},
  \bibinfo{author}{\bibfnamefont{M.}~\bibnamefont{Rich}}, \bibnamefont{and}
  \bibinfo{author}{\bibfnamefont{W.~M.} \bibnamefont{Visscher}},
  \bibinfo{journal}{Phys.\ Rev.} \textbf{\bibinfo{volume}{160}},
  \bibinfo{pages}{706} (\bibinfo{year}{1967}).

\bibitem[{\citenamefont{Evans and Morriss}(1990)}]{Evans90}
\bibinfo{author}{\bibfnamefont{D.~J.} \bibnamefont{Evans}} \bibnamefont{and}
  \bibinfo{author}{\bibfnamefont{G.~P.} \bibnamefont{Morriss}},
  \emph{\bibinfo{title}{Statistical Mechanics of Nonequilibrium Liquids}}
  (\bibinfo{publisher}{Academic Press}, \bibinfo{address}{London},
  \bibinfo{year}{1990}).

\bibitem[{\citenamefont{Evans}(1982)}]{Evans82}
\bibinfo{author}{\bibfnamefont{D.~J.} \bibnamefont{Evans}},
  \bibinfo{journal}{Phys.\ Lett.} \textbf{\bibinfo{volume}{91A}},
  \bibinfo{pages}{457} (\bibinfo{year}{1982}).

\bibitem[{\citenamefont{Fr{\"o}hlich et~al.}(1986)\citenamefont{Fr{\"o}hlich,
  Spencer, and Wayne}}]{Frohlich86}
\bibinfo{author}{\bibfnamefont{J.}~\bibnamefont{Fr{\"o}hlich}},
  \bibinfo{author}{\bibfnamefont{T.}~\bibnamefont{Spencer}}, \bibnamefont{and}
  \bibinfo{author}{\bibfnamefont{C.~E.} \bibnamefont{Wayne}},
  \bibinfo{journal}{J.\ Stat.\ Phys.} \textbf{\bibinfo{volume}{42}},
  \bibinfo{pages}{247} (\bibinfo{year}{1986}).

\bibitem[{\citenamefont{Snyder and Kirkpatrick}(1999)}]{Snyder99}
\bibinfo{author}{\bibfnamefont{K.~A.} \bibnamefont{Snyder}} \bibnamefont{and}
  \bibinfo{author}{\bibfnamefont{T.~R.} \bibnamefont{Kirkpatrick}},
  \bibinfo{journal}{Ann.\ Phys.\ (Leipzig)} \textbf{\bibinfo{volume}{8}},
  \bibinfo{pages}{SI 241} (\bibinfo{year}{1999}).

\bibitem[{\citenamefont{Cretegny et~al.}(1997)\citenamefont{Cretegny, Dauxois,
  Ruffo, and Torcini}}]{Cretegny97}
\bibinfo{author}{\bibfnamefont{T.}~\bibnamefont{Cretegny}},
  \bibinfo{author}{\bibfnamefont{T.}~\bibnamefont{Dauxois}},
  \bibinfo{author}{\bibfnamefont{S.}~\bibnamefont{Ruffo}}, \bibnamefont{and}
  \bibinfo{author}{\bibfnamefont{A.}~\bibnamefont{Torcini}},
  \emph{\bibinfo{title}{Localization and equipartition of energy in the
  beta-{FPU} chain : {C}haotic breathers}} (\bibinfo{year}{1997}),
  \eprint{cond-mat/9709204}.

\bibitem[{\citenamefont{Brillouin}(1956)}]{Brillouin56}
\bibinfo{author}{\bibfnamefont{L.}~\bibnamefont{Brillouin}},
  \emph{\bibinfo{title}{Science and Information Theory}}
  (\bibinfo{publisher}{Academic Press}, \bibinfo{address}{New York},
  \bibinfo{year}{1956}).

\bibitem[{\citenamefont{Luca and Lichtenberg}(2002)}]{Luca02}
\bibinfo{author}{\bibfnamefont{J.~D.} \bibnamefont{Luca}} \bibnamefont{and}
  \bibinfo{author}{\bibfnamefont{A.}~\bibnamefont{Lichtenberg}},
  \bibinfo{journal}{Phys.\ Rev.\ E} \textbf{\bibinfo{volume}{66}},
  \bibinfo{pages}{026206} (\bibinfo{year}{2002}).

\bibitem[{\citenamefont{Saxon and Hunter}(1949)}]{Saxon49}
\bibinfo{author}{\bibfnamefont{D.~S.} \bibnamefont{Saxon}} \bibnamefont{and}
  \bibinfo{author}{\bibfnamefont{R.~A.} \bibnamefont{Hunter}},
  \bibinfo{journal}{Philips Res.\ Rep.} \textbf{\bibinfo{volume}{4}},
  \bibinfo{pages}{81} (\bibinfo{year}{1949}).

\bibitem[{\citenamefont{Luttinger}(1951)}]{Luttinger51}
\bibinfo{author}{\bibfnamefont{J.~M.} \bibnamefont{Luttinger}},
  \bibinfo{journal}{Philips Res.\ Rep.} \textbf{\bibinfo{volume}{6}},
  \bibinfo{pages}{303} (\bibinfo{year}{1951}).

\bibitem[{\citenamefont{Hori}(1968)}]{Hori68}
\bibinfo{author}{\bibfnamefont{J.}~\bibnamefont{Hori}},
  \emph{\bibinfo{title}{Spectral Properties of Disordered Chains and Lattices}}
  (\bibinfo{publisher}{Pergamon Press}, \bibinfo{address}{Oxford},
  \bibinfo{year}{1968}).

\bibitem[{\citenamefont{Payton and Visscher}(1968)}]{Payton68}
\bibinfo{author}{\bibfnamefont{D.~N.} \bibnamefont{Payton}} \bibnamefont{and}
  \bibinfo{author}{\bibfnamefont{W.~M.} \bibnamefont{Visscher}},
  \bibinfo{journal}{Phys.\ Rev.} \textbf{\bibinfo{volume}{175}},
  \bibinfo{pages}{1201} (\bibinfo{year}{1968}).

\bibitem[{\citenamefont{Li and Wang}(2003)}]{Li03}
\bibinfo{author}{\bibfnamefont{B.}~\bibnamefont{Li}} \bibnamefont{and}
  \bibinfo{author}{\bibfnamefont{J.}~\bibnamefont{Wang}},
  \bibinfo{journal}{Phys.\ Rev.\ Lett.} \textbf{\bibinfo{volume}{91}},
  \bibinfo{pages}{044301} (\bibinfo{year}{2003}).

\end{thebibliography}

\end{document}